\documentclass{aa}
\usepackage{graphicx}
\def\ltsima{$\; \buildrel < \over \sim \;$}
\def\simlt{\lower.5ex\hbox{\ltsima}}
\def\gtsima{$\; \buildrel > \over \sim \;$}
\def\simgt{\lower.5ex\hbox{\gtsima}}
\def\sm{\mbox{M$_\odot$}}
\def\sol{\mbox{L$_\odot$}}
\def\sr{\mbox{R$_\odot$}}
\def\roph{\mbox{$\rho$~Ophiuchi}}
\def\aiso{\mbox{$\alpha_\mathrm{IR}^{7-14}$}}
\def\air{\mbox{$\alpha_\mathrm{IR}$}}
\def\aircl{\mbox{$\alpha_\mathrm{IR}^{2-14}$}}
\def\airc{\mbox{$\alpha_\mathrm{IR}^{2-7}$}}
\def\aircc{\mbox{$\alpha_\mathrm{IR}^{2-10}$}}

\def\av{\mbox{$A_V$}}
\def\ls{\mbox{$L_\star$}}

\def\lcal{\mbox{$L_\mathrm{cal}$}}
\def\lbol{\mbox{$L_\mathrm{bol}$}}
\def\lacc{\mbox{$L_\mathrm{acc}$}}

\def\ms{\mbox{$M_\star$}}

\def\mflat{\mbox{$M_\mathrm{flat}$}}
\def\mgas{\mbox{$M_\mathrm{gas}$}}
\def\mst{\mbox{$M_\mathrm{star}$}}

\def\ld{\mbox{$L_\mathrm{disk}$}}
\def\rs{\mbox{$R_\star$}}
\def\sr{\mbox{R$_\odot$}}

\def\teff{\mbox{$T_\mathrm{eff}$}}

\def\flwd{\mbox{$F_\nu^{6.7}$}}
\def\flwt{\mbox{$F_\nu^{14.3}$}}
\def\flwds{\mbox{$F_{\nu\hspace{0.1cm}\star}^{6.7}$}}
\def\flwts{\mbox{$F_{\nu\hspace{0.1cm}\star}^{14.3}$}}
\def\siglwd{\mbox{$\sigma_{6.7}$}}
\def\siglwt{\mbox{$\sigma_{14.3}$}}
\def\tint{\mbox{$t_\mathrm{int}$}}
\def\rchi{\mbox{$\chi^2\!\!_\nu$}}
\begin{document}

   \title{ISOCAM observations of the $\rho$~Ophiuchi cloud: Luminosity and 
mass functions of the pre-main sequence embedded cluster 
\thanks{This work is based on observations with
ISO, an ESA project with instruments funded by ESA Member States 
(especially the PI countries: France, Germany, the Netherlands, and
the United Kingdom) with the participation of ISAS and NASA.}\fnmsep
\thanks{Table 1 is only available in electronic form at 
the CDS via anonymous ftp to cdsarc.u-strasbg.fr (130.79.128.5) or via
http://cdsweb.u-strasbg.fr/}}

   \author{S. Bontemps\inst{1,2}
  \and  P. Andr\'e\inst{3} \and A.A. Kaas\inst{4,2} \and L. Nordh\inst{2}
  \and G. Olofsson\inst{2} \and  M. Huldtgren\inst{2} \and A. Abergel\inst{5}
  \and  J. Blommaert\inst{6} \and F. Boulanger\inst{5} \and M. Burgdorf\inst{6}
  \and  C.J. Cesarsky\inst{3} \and D. Cesarsky\inst{5} \and E. Copet\inst{7}
  \and  J. Davies\inst{8} \and E. Falgarone\inst{9} \and G. Lagache\inst{5}
  \and  T. Montmerle\inst{3} \and M. P\'erault\inst{9} \and P. Persi\inst{10}
  \and  T. Prusti\inst{6} \and J.L. Puget\inst{5} \and F. Sibille\inst{11}
      }
   \offprints{S. Bontemps, \email{bontemps@observ.u-bordeaux.fr}}

   \institute{Observatoire de Bordeaux, B.P. 89, 33270 Floirac, France\\
              email: bontemps@observ.u-bordeaux.fr
     \and
        Stockholm Observatory, 133 36 Saltsj\"obaden, Sweden
     \and
        Service d'Astrophysique, CEA Saclay, 91191 Gif-sur-Yvette, France
     \and
        ESA/ESTEC, Astrophysics Division, Netherlands
     \and
        IAS, Universit\'e Paris XI, 91405 Orsay, France
     \and
        ISO/SOC, Astrophysics Division of ESA, Villafranca, Spain
     \and
        DESPA, Obs. Paris-Meudon, 5 Pl. J. Janssen, 92195 Meudon, France
     \and
        JAC, 660 N.A'Ohoku Place, University Park, Hilo, HI 96720, USA
     \and
        ENS Radioastronomie, 24 Rue Lhomond, 75231 Paris, France
     \and
        IAS, CNR, Area di Ricerca Tor Vergata, 00133 Roma, Italy
     \and
        Observatoire de Lyon, 69230 Saint Genis Laval, France
      }

   \date{Received June 2000 / Accepted March 2001}

   \abstract{
We present the results of the first extensive 
mid-infrared (IR) imaging survey 
of the $\rho$~Ophiuchi embedded cluster, performed 
with the ISOCAM camera on board the ISO satellite.
The main $\roph$ molecular cloud L1688, as well as the two 
secondary clouds L1689N and L1689S, have been completely surveyed 
for point sources at 6.7~$\mu$m and 14.3~$\mu$m. 
A total of 425 sources are detected in $\sim 0.7$~deg$^2$, including
16 Class~I, 123 Class~II, and 77 Class~III young stellar objects (YSOs).
Essentially all of the mid-IR sources coincide with near-IR 
sources, but a large proportion of them are recognized for the first
time as YSOs. Our dual-wavelength survey allows us to identify essentially 
all the YSOs with IR excess in the 
embedded cluster
down to $F_\nu \sim $~10--15~mJy. 
It more than doubles the known population of Class~II YSOs
and represents the most complete census to date of 
newly formed stars in the $\rho$~Ophiuchi central region.
There are, however, reasons to believe that several tens of 
Class~III YSOs remain to be identified below  $L_\star \sim 0.2\ \sol$. 
The mid-IR luminosities of most ($\sim 65$\%) Class~II objects are 
consistent with emission from purely passive circumstellar disks. 
The stellar luminosity function of the complete sample of Class~II 
YSOs
is derived with a good accuracy down to $L_\star \sim 0.03\ \sol$.
It is basically flat (in logarithmic units)  
below $L_\star \sim  2\ \sol$, exhibits a possible local maximum at 
$L_\star \sim 1.5\ \sol$,
and sharply falls off at higher luminosities. 
A modeling of the luminosity function, using 
available pre-main sequence tracks and plausible star formation histories,
allows us to derive the mass distribution of the Class~II YSOs  
which arguably reflects the initial mass function (IMF) of the 
embedded cluster.
After correction for the presence of unresolved binary systems, 
we estimate that the IMF in $\roph$ is well described by a two-component 
power law 
with a low-mass index of $-0.35\pm0.25$, a high-mass index of $-1.7$
(to be compared with the Salpeter value of $-1.35$), and a break 
occurring at $\mflat=0.55\pm0.25\,\sm$. 
This IMF is flat with 
no evidence for a low-mass cutoff down to at least $\sim 0.06\,\sm$.
}
   \titlerunning{ISOCAM observations of the $\rho$~Ophiuchi cloud}
   
\maketitle

%
\section{Introduction} 

Recent observations suggest that most stars in our Galaxy and other
galaxies
form in compact clusters.
In particular, near-IR imaging surveys of nearby molecular 
cloud complexes have shown that the star formation activity 
is typically concentrated within a few 
rich clusters associated with massive dense cores which constitute 
only a small fraction of the total gas mass available
(e.g. Lada, Strom, \& Myers \cite{lada93}, Zinnecker, McCaughrean, \& 
Wilking \cite{zinnecker}). 
These embedded clusters comprise various types of young stellar
objects (YSOs) -- from still collapsing protostars to young 
main sequence stars -- but are usually dominated in number by 
low-mass pre-main sequence (PMS) stars, i.e., T~Tauri stars. 
Young clusters provide excellent laboratories for studying
the formation and early evolution of stars 
through the observational analysis 
of large, genetically homogeneous samples of embedded YSOs. 
Two key characteristics of these young stellar populations are their 
luminosity distribution and their mass spectrum,  
which give important observational constraints on the 
stellar initial mass function (IMF) in the Galaxy.
Observations of young embedded clusters can also help us understand possible 
links between parent cloud properties and the resulting stellar masses. 
They are however hampered by: (1) 
dust extinction from the parent cloud which hides most of the newly formed
stars at optical wavelengths, (2) the difficulty to recognize the nature
of individual sources 
(e.g. protostars, T~Tauri stars, or background sources), (3) the youth 
(and thus poorly known intrinsic properties) of most cluster members. 

Owing to these difficulties,  
the census of embedded YSOs provided by IRAS and 
near-IR studies is far from complete even in the nearest clouds  
(e.g. Wilking, Lada, \& Young \cite{wly}  -- hereafter WLY89). 
Thanks to its high sensitivity and good spatial resolution 
in the mid-IR, the ISOCAM camera on board ISO (Cesarsky et al. \cite{cesarsky};
Kessler et al. \cite{kessler}) 
was a powerful tool to achieve more complete surveys for YSOs in 
all major nearby star-forming regions 
(see Nordh et al. \cite{nordh1}, \cite{nordh2}; 
Olofsson et al. \cite{ohk99}; Persi et al. \cite{pmo00}). 

The nearby $\roph$ cloud is one of the most actively studied sites of 
low-mass star formation. Its central region harbors  
a rich embedded cluster with about 100 members recognized prior 
to the present work
(e.g. WLY89, Casanova et al. \cite{cmfa}).
While a dispersed population of optically visible young stars, 
associated with the Upper-Scorpius OB association, has a typical age of 
several million years (Myr) (e.g. Preibisch \& Zinnecker \cite{pz99}), the central embedded 
cluster is recognized as one of the youngest clusters known with  
an estimated age on the order of $\sim $~0.3--1~Myr (e.g. WLY89, 
Greene \& Meyer \cite{gm95}, Luhman \& Rieke \cite{lr99}). 
This young  cluster has been extensively studied at 
wavelengths ranging from the X-ray to the radio band. 
The satellites {\em Einstein} and $ROSAT$ have revealed $\sim 70$ 
highly variable X-ray sources associated with magnetically-active young stars, 
including deeply embedded protostellar sources 
(Montmerle et al. \cite{montmerle83}, Casanova et al. \cite{cmfa}, 
Grosso et al. \cite{grosso}). 
In the near-IR, the cloud has been deeply  
surveyed from the ground using large-format arrays (Greene \& Young 
\cite{gy}, Comer\'on et al. \cite{crbr}, Strom et al. \cite{sks}, Barsony et 
al. \cite{bklt}).
Unfortunately, due to difficulties in discriminating between background sources
and embedded YSOs without performing time-consuming mid-IR photometry (e.g. 
Greene et al. \cite{gwayl}) or near-IR spectroscopy
(e.g. Greene \& Lada \cite{gl96}; Luhman \& Rieke \cite{lr99}), 
these recent near-IR surveys have only partially 
increased the number of classified, recognized members. 
Finally, while only relatively poor angular resolution $IRAS$ data are 
available so far in the far-IR (e.g. WLY89), deep imaging surveys 
at an angular resolution of $10-15\arcsec$ or better exist at (sub)millimeter 
wavelengths (e.g., Motte, Andr\'e, \& Neri \cite{man} -- hereafter MAN98 --; 
Wilson et al. \cite{wilson99}).

It is thanks to the illuminating 
example of the $\roph$ embedded cluster that the now 
widely used empirical classification of YSOs was originally introduced. 
Three IR classes were initially distinguished based on the shapes of 
the observed spectral energy distributions (SEDs) between $\sim $~2~$\mu$m 
and $\sim $~25--100~$\mu$m (Lada \& Wilking \cite{lw84}, WLY89). 
Objects with rising SEDs in this wavelength range were classified as Class~I, 
sources with SEDs broader than blackbodies but decreasing longward of 
$\sim $~2~$\mu$m as Class~II, and sources with SEDs consistent with
(or only slightly broader than) reddened stellar blackbodies
as Class~III.  
These morphological SED classes are interpreted in terms of an
evolutionary sequence 
from (evolved) protostars (Class~I), to T Tauri stars with optically thick 
IR circumstellar disks (Class~II), to weak T Tauri stars with at 
most optically thin disks (Class~III) 
(Lada \cite{lada87}, Adams, Lada, \& Shu \cite{als},
Andr\'e \& Montmerle \cite{am} -- hereafter AM94). 
A fourth class (Class~0) was subsequently introduced by 
Andr\'e, Ward-Thompson, \& Barsony (\cite{awb}) 
to accommodate the discovery in the radio range 
of cold sources with large submillimeter to 
bolometric luminosity ratios 
and powerful jet-like outflows, such as 
VLA~1623 in $\rho $~Oph~A (e.g. Andr\'e et al. \cite{andre90}; Bontemps
et al. \cite{batc}). 
Class~0 objects, 
which have measured circumstellar envelope masses larger than their inferred  
central stellar masses, 
are interpreted as young protostars at the beginning of the main 
accretion phase 
(e.g. Andr\'e, Ward-Thompson, \& Barsony \cite{awb00}).
The fact that the $\roph$ central region contains at least two Class~0 
protostars (Andr\'e, Ward-Thompson, \& Barsony \cite{awb})), 
as well as numerous ($\simgt 60$) pre-stellar condensations 
(MAN98), demonstrates that it is still actively forming stars at the present
time.
 
The distance to the $\roph$ cloud is somewhat uncertain.
Usually, a value of 160~pc is adopted (e.g. Chini \cite{chini81}). 
However, recent Hipparcos results on the Upper-Scorpius OB 
association (de Zeeuw et al. \cite{dezeeuw}) 
provide a reasonably accurate estimate of $145\pm 2$~pc for the average
distance to the stars of the OB association. The 
embedded cluster is located at the inner edge of the molecular
complex on the outskirts of the OB association (e.g. de Geus \cite{degeus}), 
and not very far, in projection, from the association
center (less than $4\degr$ apart which corresponds to $\sim 10$~pc;
see Fig.~9 of de Zeeuw et al. \cite{dezeeuw}). 
In this paper, we therefore 
adopt a distance of $d = 140\pm 10$~pc
for the $\roph$ IR cluster which corresponds to a distance modulus 
$5\times$log$_{10}(d/10\rm pc) = 5.73\pm0.15$.

The layout of the paper is as follows. 
Sect.~2 gives observational details (\S~2.2) 
and describes the way the data have been reduced to obtain mid-IR 
images
and extract point-sources (\S~2.3) along with the photometric uncertainties 
and the sensitivity levels (\S~2.4). In Sect.~3, the identification of
detected sources is discussed (\S~3.1) and the selection of a new population
of 123 Class~II YSOs, as well as 16 Class~I and 77 Class~III YSOs,
is described (\S~3.2-3.5). We then derive 
luminosities for these YSOs and build the corresponding
luminosity functions in Sect.~4. 
In Sect.~5, we model the 
luminosity functions for Class~II and Class~III YSOs in terms of the underlying
mass function and star formation history. 
In Sect.~6, we discuss the resulting 
constraints on the IMF of the $\roph$ cloud down to $\sim 0.06\,\sm$, 
(\S~6.1-6.2)
as well as related implications (\S~6.3-6.5).
\section{Observations and data reduction}

\subsection{Region surveyed by ISOCAM}

\begin{figure*}[t]
\includegraphics[height=17.8cm,angle=-90]{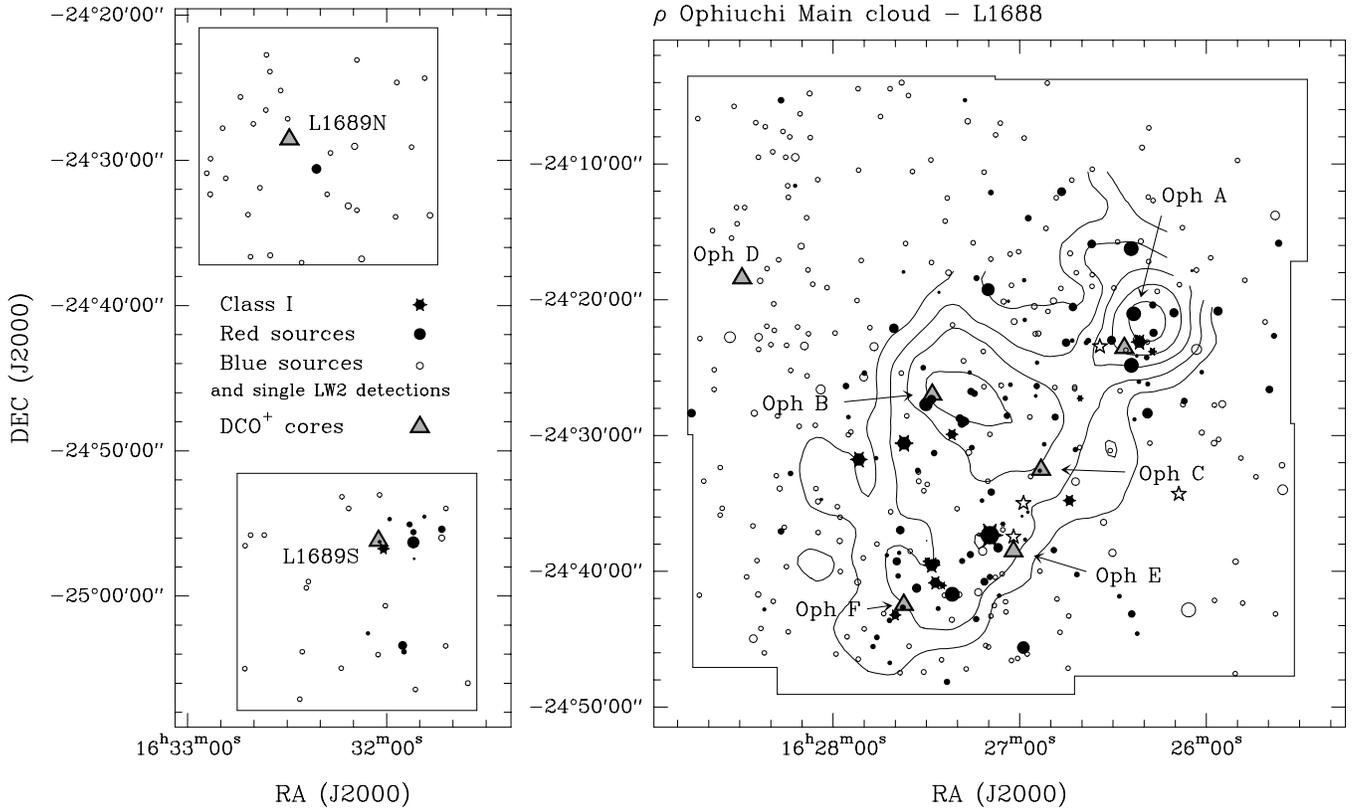}
\label{map}
\caption[]{
Sky map of the $\roph$ clouds L1688, L1689N, and L1689S 
showing the spatial distribution
of the ISOCAM sources. The filled symbols correspond to ``red'' YSOs 
(i.e. Class~I and Class~II YSOs, see Sect.~3), while 
the open circles mark the ``blue'' sources. 
The surface areas of these symbols are 
proportional to the 6.7$\mu$m flux densities.
The open stars mark four young early-type stars (SR3, S1, WL22, WL16 from right to 
left; see Sect.~3.1).
The contours show the L1688 molecular cloud as mapped 
in CS(2-1) by Liseau et al. (1995).
The triangles refer to the peaks of the DCO$^+$ dense cores of 
Loren et al. (1990). The figure emphasizes the presence of four 
sub-clusters Oph~A, B, EF (between Oph~E and Oph~F), and L1689S 
(named after the associated dense cores). 
}
\end{figure*}

The survey encompasses  
the $\rho$ Ophiuchi central region associated with the 
prominent dark cloud L1688, as well as 
the two subsidiary sites L1689N and L1689S.
The L1688 field is a $\sim 45\arcmin \times 45\arcmin$ square, 
while the L1689N and L1689S fields each cover an area of 
$16.5\arcmin \times 16\arcmin$ (see Fig.~1).
Most previously known members of the $\rho$~Ophiuchi cluster lie within 
these fields. In particular, this is the case for 
94 of a total of 113 recognized 
members from WLY89, AM94, 
Greene et al. (\cite{gwayl}), and MAN98. 
The known young stars which lie outside the boundaries of our survey 
are mostly optically visible, weak-line or post T Tauri stars (belonging to 
Class~III) spread over a large area on the outskirts of the 
molecular complex (e.g. Mart\'\i n et al. \cite{martin98}). 

\subsection{Observational details}

The mapping was performed in the raster mode of ISOCAM in which 
the mid-IR $32\times32$ pixel array 
imaged the sky at consecutive positions along 
a series of scans parallel to the right-ascension axis. 
The offset between consecutive array positions along each scan 
($\Delta \alpha$) 
was 15 pixels, while the offset between 
two scans ($\Delta \delta$) was 26 pixels. 
Each set of scans was then co-added and combined into a single raster image.
The final image of the L1688 field (Fig.~1 and Abergel et al. 1996)
actually results from the combination of six separate rasters.
A pixel field of view of 6$\arcsec$ was used for four of these rasters, 
but smaller 3$\arcsec$ pixels were employed for the other two rasters in order 
to avoid saturating the array on the brightest sources of the 
cluster.
The L1689N and L1689S fields were imaged with one raster each
using 6$\arcsec$ pixels.

In order to avoid saturation, the individual readout time for the L1688
rasters was set to $\tint = 0.28\,$s. About 55 of these readouts
(i.e. an integration time of $\sim 15\,$s) were performed 
per sky position. Thanks to the half-frame overlap between subsequent 
individual images, each sky position was observed twice, 
yielding an effective total integration time of $\sim 30\,$s. 
For L1689N and L1689S it was possible to use $\tint = 2.1\,$s, and about 15 
readouts were performed per sky position with the same half-frame overlap, 
giving an integration time per sky position of $\sim 60\,$s.
A total of 1104 individual images were necessary
to mosaic the L1688 field, and an additional 60 images each were used to map 
L1689N and L1689S.
 
All three fields were mapped in two broad--band filters of 
ISOCAM: LW2 (5--8.5\,$\mu$m) and LW3 (12--18\,$\mu$m). 
These filters are approximately centered on two minima of the interstellar 
extinction curve and are situated apart from the silicate absorption 
bands (at roughly 10 and 18~$\mu$m). 
However, they include most of the Unidentified Infrared Bands
(UIBs, likely due to PAH-like molecules) 
which constitute a major source of background emission toward
star-forming clouds (e.g. Bernard et al. \cite{bernard}, 
Boulanger et al. \cite{boulanger}). 
The ISOCAM central wavelengths adopted here for LW2 and LW3 are 
6.7~$\mu$m and 14.3~$\mu$m respectively.

\subsection{Image processing, source extraction, and photometry}

Each raster consists of a temporal series of individual integration
frames (i.e. of $32\times32$ pixel images) which was reduced using 
the CAM Interactive Analysis software (CIA)\footnote{
CIA is a joint development by the ESA Astrophysics Division and
the ISOCAM consortium led by the ISOCAM PI, C.~Cesarsky, Direction
des Sciences de la Mati\`ere, C.E.A., France.}.
We have subtracted the best dark current from the ISOCAM calibration 
library, and as a second step we improved it with a second order correction
using a FFT thresholding method (Starck et al. \cite{starck2}). 
Cosmic-ray hits were detected and masked using the
multi-resolution median transform algorithm (Starck et al. \cite{starck1}). 
The transients in the time history of each pixel due to detector 
memory effects were corrected with the inversion method described 
in Abergel et al. (\cite{abergel}). 
The images were then flat-fielded with a flat image obtained from
the observations themselves.
Since these various corrections applied to the images are not 
perfect, the extraction of faint sources from the images is 
a difficult task. We have developed an interactive IDL
point-source detection and photometry program for raster observations
which works in the CIA environment. This program helps to discriminate between 
astronomical sources and remaining low-level glitches
or ghosts due to strong transients (see also Nordh et al. \cite{nordh1};
Kaas et al. \cite{kaas01}). 
The fluxes of the detected sources were estimated from the series of
flux measurements made in the individual images (usually 2 to 
4 individual images cover each source)
which were obtained from classical aperture
photometry.
The emission was integrated in a sky aperture,
the background emission subtracted, and finally an appropriate 
aperture correction was applied 
based on observed point-spread functions available in the 
ISOCAM calibration library.
In practice, the radius of the aperture used was 9$\arcsec$
(i.e., 3 and 1.5 pixels for a pixel size of $3\arcsec$ and 
6$\arcsec$, respectively). For the weakest sources, however, we reduced
the aperture radius to 4.5$\arcsec$ 
(i.e. 1.5 pixels for a pixel size of $3\arcsec$), in order to improve the 
signal-to-noise ratio.
Finally, we applied the following conversion factors: 
$2.33$ and 
$1.97$~ADU/gain/s/mJy for LW2 and LW3 respectively (from  
in-orbit latest calibration - Blommaert \cite{blommaert}). These 
calibration factors are strictly valid only for sources
with a flat SED  ($F_\nu \sim \nu^{-1}$). 
Here, a small but significant ($\simgt 1$~\%) color correction needs to be 
applied to the bluest sources, recognized
as Class~III YSOs in Sect.~3 below. For these sources, 
the conversion factors quoted above were divided by 1.05 for LW2 
and 1.02 for LW3 to account for the color effect.
The 212 ISOCAM sources recognized as cluster members (see Sect.~3 below) 
are listed in Table~1 (available only in electronic form at 
http://cdsweb.u-strasbg.fr/) with their J2000 coordinates, their flux 
densities and associated rms uncertainties (see Sect. 2.4), as well as 
the corresponding near-IR identifications.

\subsection{Photometric uncertainties and point-source sensitivity}

The uncertainties on the final photometric measurements result from 
systematic errors due to uncertainties in the 
absolute calibration and the aperture correction factors, and 
from random errors associated with the flat-fielding noise, the statistical 
noise in the raw data, the noise due to remaining low-level glitches, 
and the imperfect correction for the transient behavior of the detectors.
The in-orbit absolute calibration has been verified to be 
correct to within 5~\% (Blommaert \cite{blommaert}), and we estimate that
the maximum systematic error on the aperture correction is  
$\sim 10\,$\% 
(by comparing theoretical and observed point-spread functions). The
maximum systematic error on our photometry is thus $\sim 15\,$\%.
The magnitudes of the random errors were directly estimated from the 
data by measuring both a ``temporal'' noise (noise in the temporal
sequence of individual integrations) and a ``spatial'' noise (due to 
imperfect flat-fielding and/or spatial structures in the local
mid-IR background emission) for each source in the automatic 
detection procedure. 
The temporal noise was computed as the
standard deviation of the individual aperture measurements divided by the 
square root of the number of measurements. 
The spatial noise was estimated  
as the standard deviation around the mean background
(linear combination of the median and the mean 
of the pixels optimized for the source flux estimates) 
in the immediate vicinity of each source.

\begin{figure}[ht]
\centering
\includegraphics[width=8.7cm]{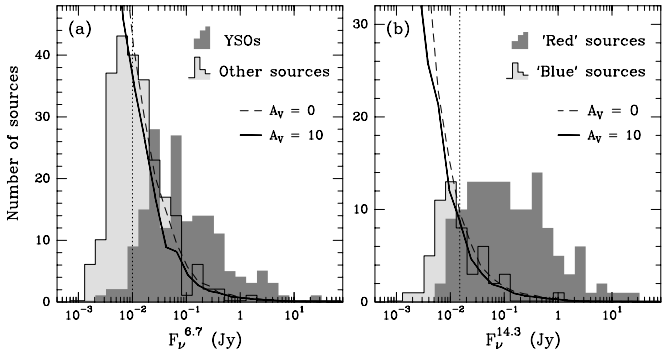}
\label{histo}
\caption[]{
{\bf (a)} Distribution of the $\flwd$ flux densities 
for the 425 sources detected by ISOCAM. 
The dark histogram shows the flux distribution of the cluster members  
(i.e., YSOs) discussed in Sect.~3, while the light histogram corresponds to the 
remaining sources which are likely dominated by background Galactic objects.
{\bf (b)} Same as (a) for the $\flwt$ flux densities. The dark histogram shows the 
flux distribution for the ``red'' ISOCAM sources (i.e., Class~I and II YSOs), 
while the light histogram comprises only the ``blue'' ISOCAM sources 
{\it not} associated with recognized YSOs (see Sect.~3 and Fig.~3).
The adopted completeness levels equal to 10 and $15\,$mJy for $\flwd$ and $\flwt$ 
respectively are shown as vertical dotted lines.
The solid and dashed curves show the expected number 
of Galactic sources in each bin according to the model of 
Wainscoat et al. (1992) for $A_V  = 0 $ and $A_V  = 10 $ respectively.
}
\end{figure}

The sensitivity limit of the survey was estimated by 
calculating the average value of the quadratic sum of 
the temporal and spatial noises measured on the weakest detected sources. 
The total rms flux uncertainty found in this way, 
$\sigma_{tot} = (\sigma_{temp}^2 +\sigma_{spat}^2)^{1/2} $,  
is $\siglwd = 2.2$~mJy at 6.7~$\mu$m 
and $\siglwt = 4.1$~mJy at 14.3~$\mu$m, 
$\sim $~75~\% of which is due to the spatial noise component.
The large contribution
of the spatial noise originates in the highly structured   
diffuse mid-IR emission from the ambient molecular cloud itself 
(see Abergel et al. \cite{abergel}).
Figure~2 displays the distributions of fluxes at 6.7~$\mu$m and 14.3~$\mu$m 
for all the detected ISOCAM sources. We used the Wainscoat et al. 
(\cite{wainscoat}) 
Galactic model of the mid-IR point source sky to estimate the expected 
number of foreground and background sources up to a distance of 20$\,$kpc. 
The model predictions are shown by solid and dashed curves in Fig.~2
for cloud extinctions of $A_V  = 0 $ and $A_V  = 10 $, respectively
(see Kaas et al. \cite{kaas01} for more details). 
It can be seen that the flux histograms of the ISOCAM sources 
not associated with YSOs (light shading in Fig.~2) are remarkably 
similar in shape to the model distributions down to $\sim 6$~mJy at 6.7~$\mu$m 
and $\sim 10$~mJy at 14.3~$\mu$m. These flux densities can be used to   
estimate the completeness level of our observations which is not uniform 
over the spatial extent of the survey. 
The histograms with light (grey) shading in Fig.~2 
are dominated by background sources preferentially located in  
low-noise regions (i.e., outside the crowded central part of L1688), where 
the total rms flux uncertainty 
is $\sim 2.0$~mJy at 6.7~$\mu$m and $\sim 3.5$~mJy at 14.3~$\mu$m. 
The effective completeness level in these regions 
is thus $\sim 3\,\sigma_{tot}$, where $\sigma_{tot} $
is the total\footnote{
The detection level is more directly related to the
temporal noise than to the spatial noise. 
In terms of $\sigma_{temp}$, the effective completeness level is  
$\sim 12\,\sigma_{temp}$.}
flux uncertainty (see above).
However, most of the YSOs are located in regions where the noise is 
somewhat larger. The largest rms noise is reached in the Oph~A core area 
(see Fig.~1), where $\sigma_{tot} \approx 3.4\,$mJy at 6.7~$\mu$m and 
$\sigma_{tot} \approx 5.0\,$mJy at 14.3~$\mu$m. 
Therefore, we conservatively estimate the completeness levels of
the global ISOCAM survey to be $\sim 10$~mJy at 6.7~$\mu$m and 
$\sim 15$~mJy at 14.3~$\mu$m.

Finally, we note that the $A_V  = 10 $ model curve in Fig.~2b
accounts for essentially all the ``blue'' sources detected  
at 14.3~$\mu$m and not associated with known YSOs. At 6.7~$\mu$m, 
the predictions of the Wainscoat et al. model suggest that there 
might still be a slight excess of $\sim 30$ unidentified 
sources belonging to the cloud (Fig.~2a). 

\section{Identification and nature of the mid-IR sources}

\subsection{Statistics of detections}

Within the $\sim 0.7$~square degree imaged by ISOCAM,
a total of 425 sources has been identified, among which 211 are seen at both
6.7~$\mu$m and 14.3~$\mu$m. The spatial distribution of the  
sources is shown in Fig.~1, 
where the ``red'' sources  
[those with log$_{10}$($\flwt/\flwd) > - 0.2 $, see below] 
are indicated as filled circles. 
These ``red'' sources appear to be clustered into four main groupings: three
sub-clusters in L1688, i.e., Oph~A (West), Oph~B (North-East), 
Oph~EF (South) (see also Strom, Kepner, \& Strom \cite{sks}), 
as well as a new sub-cluster in L1689S.

Four bright embedded stars (S1, SR3, WL16, WL22) are spatially resolved by 
ISOCAM in both filters.
Their extended mid-IR emission is most likely due to 
PAH-like molecules excited by relatively strong far-ultraviolet (FUV)
radiation fields\footnote{
S1 and SR3 are optically-visible stars of spectral types 
B3 and B7, respectively, which clearly emit enough UV photons 
to excite PAHs. The 
other two objects, WL16 and WL22, show PAH features in their
mid-IR spectra (e.g. Moore et al. \cite{moore}), and may be 
young embedded early-type (i.e., early A or late B) stars.}.
These bright sources are displayed as open star symbols in Fig.~1 
(SR3, S1, WL22, WL16 from right to left).

A total of 89 previously classified YSOs lie within the area of the 
present survey: 
2 Class~0, 69 Class~I/II, and 18 Class~III YSOs.
ISOCAM detected 84 of these 89 YSOs (94~\%): 
97~\% of the Class~I/IIs, 94~\% of the Class~IIIs, 
and none of the Class~0s.
The two undetected Class~I/IIs (GY256, GY257), and the undetected 
Class~III (IRS50), are located very close ($\sim $~10\arcsec--15\arcsec) 
to bright sources (WL6 and IRS48, respectively), which  
may account for their non-detection. 
The two Class~0 objects (VLA~1623 and IRAS~16293$-$2422) are 
deeply embedded within massive, cold circumstellar envelopes 
which are probably opaque at 
6.7~$\mu$m and 14.3~$\mu$m (see Andr\'e, Ward-Thompson, 
\& Barsony \cite{awb})) 
and too weak to be detected by ISOCAM. 

\subsection{A population of new YSOs with mid-IR excess}

\begin{figure*}[ht]
\centering
\includegraphics[width=13.5cm]{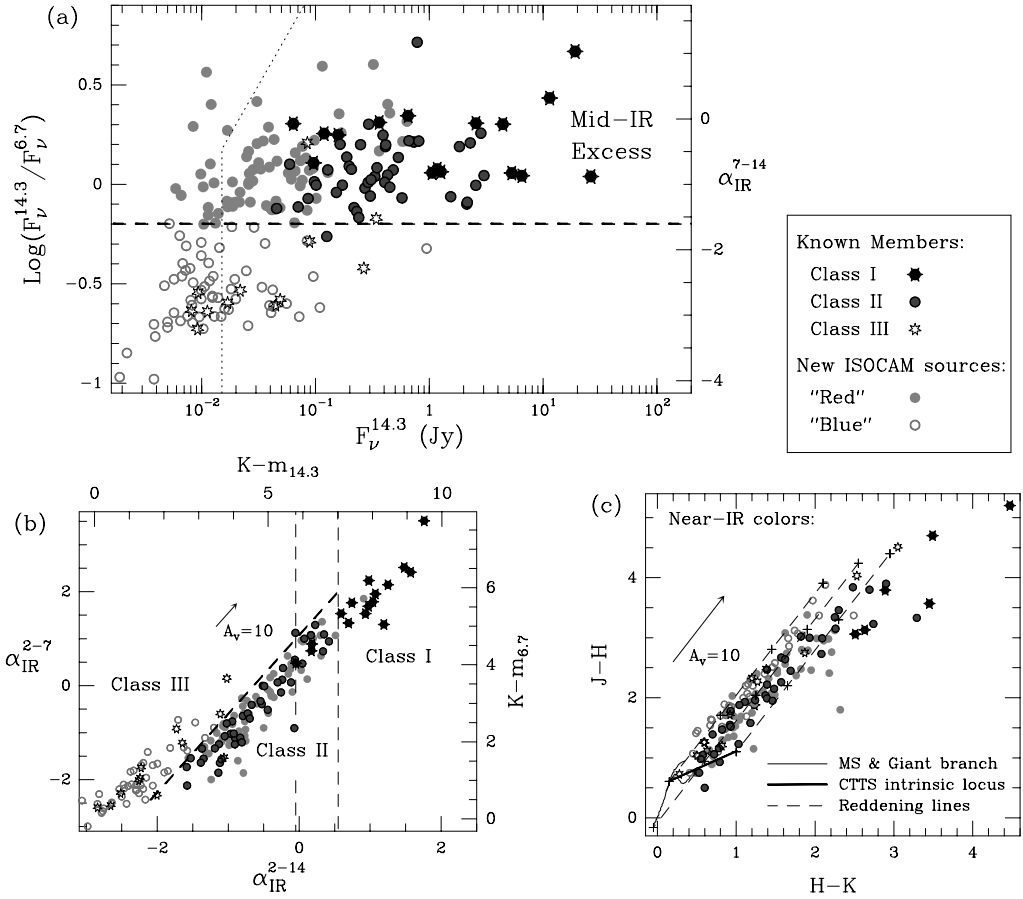}
\label{color}
\caption[]{
{\bf (a)} Logarithmic ratio of the ISOCAM fluxes, log$_{10}$($\flwt$/$\flwd$), 
versus $\flwt$ displayed for the 207 ISOCAM sources seen 
in both filters (excluding the 4 early-type young stars). 
The purely ISOCAM color $\flwt$/$\flwd$ 
is converted into an IR spectral index, 
$\aiso$, on the right axis, and can be used to identify 
sources with mid-IR excesses (above the dashed line).
The completeness limit derived in Sect.~2.4 is displayed as a dotted
line.
{\bf (b)} Mid-IR color-color diagram expressed as a 
spectral index diagram $\airc$ versus $\aircl$ for 
175 ISOCAM sources also detected in the K band (m$_{6.7}$ and m$_{14.3}$
on the right and upper axes are calculated as in Persi et al. \cite{pmo00}). 
The inclined heavy dashed line marks the 
limit to the right of which there is significant  
mid-IR excess as in (a). The two vertical dashed lines mark 
the formal boundary between Class~II, flat-spectrum, and Class~I
sources (see text for details).
{\bf (c)} Classical near-IR color-color diagram, J--H versus H--K,
for the 152 ISOCAM sources with JHK photometry.
}
\end{figure*}

As can be seen in Fig.~3, the mid-IR regime is ideal to detect and 
characterize the excess emission due to circumstellar 
disks around young stars. 
In Fig.~3a,
the 207 sources detected at 
6.7~$\mu$m and 14.3~$\mu$m (excluding S1, SR3, WL16, WL22) are shown  
in a diagram
which displays the logarithmic flux ratio log$_{10}$($\flwt/\flwd$)
against the mid-IR flux
$\flwt$. On the right axis of the diagram, 
the flux ratio $\flwt/\flwd$ has been 
converted into a classical IR spectral index, 
$\air =$~d$\log_{10}(\lambda F_\lambda)/$d$\log_{10}(\lambda)$ (e.g. WLY89), 
calculated between 6.7~$\mu$m and 14.3~$\mu$m, i.e., $\aiso$.
In this diagram, two groups of sources can clearly be distinguished. 
The lower ratio group has log$_{10}$($\flwt/\flwd) \sim -0.7$, i.e., 
a spectral index on the order of 
$\aiso \sim -3.0$, which is the 
value expected for simple photospheric blackbody emission in 
the Rayleigh-Jeans regime. 
The dispersion around $\aiso \sim -3.0$ is obviously larger for 
weaker sources. This is mainly due to increasing photometric uncertainty 
with decreasing flux.
The higher ratio group consists of ``red'' sources 
defined by log$_{10}$($\flwt/\flwd$)~$> -0.2$ (cf. Nordh et al. \cite{nordh1}) .
Most of these actually have $-0.2 \leq$~log$_{10}$($\flwt/\flwd) \leq 0.3$, 
i.e., $-1.6 \simlt \aiso \simlt 0$, which is typical of  
classical T~Tauri stars (hereafter CTTS). This range of $\aiso $   
roughly delineates the domain of Class~II YSOs
(e.g. Adams et al. 1987, AM94, Greene et al. 1994), and 
is usually interpreted in terms of 
optically thick circumstellar disk emission (e.g. Lada \& Adams \cite{la92}).  
In Fig.~3,
the new ISOCAM sources are distinguished from 
the previously known cluster members by different symbols. 
One can see in Fig.~3a that all the previously known 
Class~II sources but one lie above the dividing line for red sources, 
while all the previously known Class~III sources but two lie below it.

The object WL19, classified as a Class~I YSO by WLY89 and as a 
reddened Class~II by AM94 (see also Lada \& Wilking \cite{lw84}), 
is here found to be a ``blue'' source in the mid-IR range 
($\aiso=-1.8\pm0.2$). This object may 
correspond to a luminous Class~III star 
located behind the cloud (see also Comer\'on et al. \cite{crbr}). 
Although GY12 formally has a ``red'' mid-IR spectral 
index ($\aiso=-0.5\pm0.5$) here,
we still consider it as a Class~III object (cf. Greene et al. \cite{gwayl}). 
(The mid-IR color is highly uncertain since GY12 
is only marginaly resolved at 14.3~$\mu$m 
from its bright Class~I neighbor GSS30.)
Finally the Class~III source DoAr21 has a borderline mid-IR 
spectral index ($\aiso=-1.5\pm0.1$) but is kept as a 
Class~III object since its color between 2.2~$\mu$m and 
14.3~$\mu$m corresponds to $\aircl=-2.0\pm0.1$.

A total of 71 sources are identified for the first time as mid-IR excess 
objects in Fig.~3a.
These new ``red'' sources are most likely 
all embedded YSOs, i.e., members of the $\roph$ cluster.
Since dust extinction is roughly the same at 
6.7~$\mu$m and 14.3~$\mu$m (Rieke \& lebofsky \cite{rl}; 
Lutz \cite{lutz}), any background source should be 
intrinsically red in order to contaminate the sample of red YSOs. 
Based on the Galactic model by Wainscoat et al. (\cite{wainscoat}), 
the vast majority 
of background objects should appear ``blue'' (cf. Fig.~2).
Seven background giant stars (GY45, GY65, VSSG6, GY232, GY351,
GY411, and GY453 -- cf. Luhman \& Rieke \cite{lr99}), 
and a known foreground dwarf (HD148352 -- Garrison \cite{garrison}) 
are detected, which all have 
blue mid-IR colors ($\aiso = -2.7, -2.0, -3.1, -3.3, -2.9, -3.2, -3.1$,
and $-3.1$, respectively).

The newly identified cloud members nicely extend the previously 
known Class~I/II population toward low IR fluxes. 
While previous studies could only identify sources with 
$\flwt \simgt 100$~mJy,
the present census is complete for objects down 
to $\flwt \sim 15$~mJy.
Altogether, a sample of 139 Class~I/Class~II YSOs 
is identified, of which 71 are new members. 
The present survey has thus allowed us 
to more than double the number of recognized YSOs with 
circumstellar IR excess in the $\roph$ cloud.

The 139 red YSOs are listed in Table~2 (Class~I YSOs) 
and Table~3 (Class~II YSOs) by decreasing order of $\aircl $. 
Twelve of them (last entries of Table~3) are completely new sources 
with respect to published IR surveys.

\subsection{Class~I versus Class~II YSOs}

In Fig.~3, 
the YSOs classified as Class~I and Class~II by AM94
and Greene et al. (\cite{gwayl}) are shown as filled stars and filled 
circles, respectively. 
In the log$_{10}$($\flwt$/$\flwd$) versus $\flwt$ diagram, there is no 
clear color gap between the two classes of objects, 
even though the Class~I YSOs tend to lie in the upper part 
of the ``red'' group (see Fig.~3a).
Since extinction has a negligible effect on the mid-IR ratio
$\flwt$/$\flwd$, 
this suggests that Class~I and Class~II objects have fairly similar 
intrinsic colors between 6.7~$\mu$m and 14.3~$\mu$m. 

The classical IR spectral index
calculated from 2~$\mu$m to 10~$\mu$m (or 25~$\mu$m) 
(e.g. Lada \& Wilking \cite{lw84} and WLY89) appears to provide a better way of 
discriminating between envelope-dominated Class~I YSOs 
and disk-dominated Class~II sources (see Fig.~3b).  
In particular, millimeter continuum mapping of optically thin circumstellar 
dust emission confirms that, apart from a few important 
exceptions (e.g., WL22, WL16, WL17, IRS37, IRS47), 
the $\roph$ objects selected on the basis of 
$\aircc \simgt 0$ are indeed Class~I protostars surrounded by spheroidal 
envelopes (AM94, MAN98).\\
As expected, the previously known Class~I YSOs are 
well concentrated in the upper-right part of the $\airc$ versus $\aircl$
diagram of Fig.~3b (where $\airc$ and $\aircl$ are close to the 
$\aircc$ index used in previous studies -- 
e.g., AM94 and Greene et al. \cite{gwayl}).
There is also a hint of two gaps in this diagram 
at $\aircl \sim 0.5$ 
and $\aircl \sim 0.0$, 
which roughly bracket the regime of flat-spectrum sources as defined by 
Greene et al. (\cite{gwayl}). 
These may represent a distinct population of  
transition objects between Class~I and Class~II 
(e.g. Calvet et al. \cite{calvet94}).\\
Here, we thus consider sources with $\aircl > 0.55$ as Class~I YSOs
(Table~2), sources with $-0.05 < \aircl < 0.55 $ as candidate 
flat-spectrum objects (see Table~3), and sources with $\aircl < 0.55$ 
as Class~II YSOs (Table~3). 
These limiting indices are displayed in Fig.~3b. In the
following, the candidate flat-spectrum sources will be treated as  
Class~II YSOs.

\subsection{Mid-IR excess versus near-IR excess}

Most of the ISOCAM sources (e.g. 90~\% of the Class~II sources) 
were also detected in the 
near-IR (JHK) survey of Barsony et al. (\cite{bklt}). 
Comparison between Fig.~3b and Fig.~3c 
illustrates the advantage 
of mid-IR measurements for selecting sources with 
intrinsic circumstellar IR excesses. 
While the red and blue groups of Fig.~3a are well separated in 
the mid-IR diagram of Fig.~3b, 
they blend together in the near-IR diagram of Fig.~3c.

Figure~3c also shows 
that most of the ISOCAM-selected 
YSOs lie within the reddening band associated with the 
intrinsic locus of CTTSs as derived by Meyer et al. (\cite{meyer97}) in 
Taurus. The few exceptions, which lie to the right of the reddening band, 
correspond to Class~I and flat-spectrum YSOs.

\subsection{The Class~III YSO census in $\roph$ }

While Class~I/II YSOs are easily recognized in the mid-IR 
range thanks to their strong IR excesses, Class~III objects are difficult
to identify without
deep X-ray and/or radio centimeter continuum observations. 
We have used the ROSAT X-ray surveys of Casanova et al. (\cite{cmfa}) and 
Grosso et al. (\cite{grosso}), along with the VLA radio surveys by, e.g., 
Andr\'e, Montmerle, \& Feigelson (\cite{andre87}) and Stine et al. 
(\cite{stine88})  
to build up a sample of bona-fide Class~III YSOs covered by the present
survey. 
With the additional Class~III candidate WL19 (see Sect.~3.2 above),
there are 38 such YSOs which are listed in order of decreasing 
$\aircl$ in Table~4. 

This Class~III sample is unfortunately not as complete as the Class~I 
and Class~II samples discussed above.   
According to Grosso et al. (\cite{grosso}), the number of Class~IIIs 
may be roughly 
as large as the number of Class~IIs: above their typical X-ray 
detection limit of $L_X \sim 3 \times 10^{29}$~erg~s$^{-1}$ 
(corresponding to $L_\star \sim 0.3\, L_\odot$ -- see Fig.~7 of Grosso et al.), 
they found a Class~III/Class~II number ratio 
of 19/22 in the $ROSAT$-$HRI$/$ISO$-$ISOCAM$ overlapping survey area. 
If this ratio is representative of the 
complete population of young stars in $\roph$, the total number of Class~IIs
found here (123 objects) suggests that as many as $106$~Class~IIIs may be 
present in the cluster down to $\ls \sim 0.03\,\sol$ (our completeness level
for Class~IIs, see Sect.~4.4).  
A total of 38 Class~IIIs are already known within the ISOCAM
survey area, so that $\sim 68$ 
unknown Class~IIIs may remain to be found.
Since it was noted in Sect.~2.4 that 
$\sim 30$ sources detected at 6.7~$\mu$m 
might be unidentified cluster members, 
about half of the missing Class~IIIs 
may have been actually seen by ISOCAM. 

We also note that a large proportion (80/123) of the Class~II sources 
are closely associated with the densest part of L1688 (see Fig.~1). 
Assuming the same proportion applies to Class~IIIs, we would 
expect $\sim 44$ unknown Class~III sources to be located within  
the CS contours of Fig.~1. A total of 
39 unclassified ISOCAM sources (also detected by Barsony et al. \cite{bklt})
lie within these CS contours where the number
of detected background stars should be small due to high cloud extinction.
Most of these 39 sources might thus be yet unidentified 
Class~III YSOs. These candidate Class~III sources are listed in order of
decreasing $\aircl$ in Table~5.

\section{Luminosity estimates and luminosity function}

Most of the new YSOs identified by ISOCAM 
are weak IR sources which were not detected by IRAS  
and were not observed in previous ground-based
mid-IR surveys (dedicated to bright near-IR sources). 
They likely correspond to low-luminosity, low-mass young stars.
In Sect.~4.1 below, we derive stellar luminosity estimates
for Class~II and Class~III objects 
using published near-IR photometry from Barsony et al. (1997).
In Sect.~4.2, we provide mid-IR estimates of the disk luminosities, $\ld$, 
for Class~II YSOs.
Finally, calorimetric estimates of the bolometric
luminosities, $\lbol$, for Class~I YSOs are calculated in Sect.~4.3.
The luminosity function  
of the $\roph$ embedded cluster is then assembled in Sect.~4.4.

\subsection{Dereddened J-band stellar luminosities for Class~II and 
Class~III sources}

The J-band flux provides a good tracer of the stellar luminosity
for late-type PMS stars (i.e., T~Tauri stars) because the J-band is
close to the maximum of the photospheric energy distribution
for such cool
stars. It is also a good compromise between bands too much affected by
interstellar extinction at short wavelengths
(very few $\roph$ YSOs have been detected in the V, R, or I bands), 
and the H, K and mid-IR bands which are contaminated by 
intrinsic excesses.
Greene et al. (\cite{gwayl}) showed that there is a good 
correlation between the dereddened J-band flux and the stellar luminosity  
derived by other methods. They pointed out that in $\roph$ 
this correlation is roughly consistent with a more theoretically
based correlation expected for 1-Myr old PMS stars following the 
D'Antona \& Mazzitelli (\cite{dm94}) evolutionary tracks. 
More recently, Strom et al. (\cite{sks}) and Kenyon \& Hartmann (\cite{kh95}) 
used the same model PMS tracks to directly convert dereddened 
J-band fluxes into stellar masses. We adopt a similar approach here.

\subsubsection{Extinction estimates from near-IR colors}

The main difficulty and source of uncertainty with this method 
is due to the foreground extinction affecting the J-band fluxes.
One must estimate the interstellar extinction toward each source 
in order to correct the observed J-band fluxes.
We have used the observed near-IR colors to estimate 
the J-band extinction.
The $(J-H)$ color excess is most suitable for this purpose (e.g. Greene et al.
\cite{gwayl}) since the dispersion in the intrinsic $(J-H)$ colors of 
CTTSs is small and observationally well 
determined 
(cf. Strom et al. \cite{strom89}, Meyer et al. \cite{meyer97}).

The reddening law quoted by Cohen et al. (\cite{cohen}),
which is determined for the standard CIT system, should be applicable
to the JHK photometry of Barsony et al. (\cite{bklt}).
We have thus used: 

\begin{equation}
A_V  =  9.09\times [(J-H)-(J-H)_0] $$ 
\end{equation}

\noindent
when the $(J-H)$ color is available, and  

\noindent
\begin{equation}
A_V  =  15.4\times [(H-K)-(H-K)_0] 
\end{equation}

\noindent
otherwise, where we have adopted 
$(J-H)_{0} = 0.85$, and $(H-K)_{0} = 0.55$ from the CTTS results of  
Meyer et al. (\cite{meyer97}).
The dereddened $J_{0}$ magnitude is then obtained as
$J_{0} = J-0.265\times A_V$, or $H_{0} = H-0.155\times A_V$ 
for those stars not detected in J (with $A_V$ estimated from H--K).
The dereddened $J_{0}$ (or $H_0$) can then be converted into an absolute 
J-band (H-band) magnitude, $M_J$ ($M_H$),
via the distance modulus of the $\rho$~Ophiuchi cloud equal to
5.73 (for $d = 140$~pc).

The uncertainties on $M_J$ and $M_H$ result from 
the typical uncertainties on the J, H, K
magnitudes and on the intrinsic
colors $(J-H)_{0} $ and $(H-K)_{0}$. With $\sigma(J)=0.1$, 
$\sigma(H)\sim\sigma(K)=0.05$ (Barsony et al. \cite{bklt}), and 
$\sigma$({\small $(J-H)_{0}$})$\,= 0.1$,
$\sigma$({\small $(H-K)_{0}$})$\,= 0.2$ (Meyer et al. \cite{meyer97}), 
we obtain the following typical uncertainties: 
$\sigma$({\small $M_J$})~$= 0.39$~mag, and 
$\sigma$({\small $M_H$})~$= 0.60$~mag. In addition, the uncertainty on the 
cluster distance ($140 \pm 10$~pc) induces a maximum systematic error 
of $\pm 0.15$~mag on $M_J$ and $M_H$.

\subsubsection{Relationship between $M_J$ and $\ls$ for young T~Tauri stars}

The absolute J-band magnitude $M_J$ can be directly converted into a stellar 
luminosity $\ls$ if the effective stellar temperature $\teff$ is known:  
$\log_{10}(\ls)=1.89-0.4\times(M_{J}+BC_{J}(\teff))$, where BC$_{J}$ 
is the bolometric correction for the J band depending only on $\teff$.
Pre-main sequence objects in the mass range 
$0.1 \simlt M_\star \simlt 2.5\, M_\odot $ 
are cool sub-giant stars with typical photospheric temperatures 
$\teff \sim $~2500--5500~K (e.g. Greene \& Meyer \cite{gm95}, 
D'Antona \& Mazzitelli \cite{dm94}). 
In this temperature range (0.34~dex wide),  
the photospheric blackbody peaks close 
to the J band (1.2$\mu$m), so that the J--band bolometric correction 
spans only a limited range, $1 \simlt BC_{J} \simlt 2$, corresponding 
to a total shift in luminosity of only 0.4 dex. 
Therefore, if we use a (geometrical) average value for the 
effective temperature, $\teff^0 \sim 3700$~K, 
we should not make an error larger than 
$\pm 0.2$~dex on $\log_{10}(\ls)$.

\begin{figure}[ht]
\centering
\includegraphics[width=7.2cm]{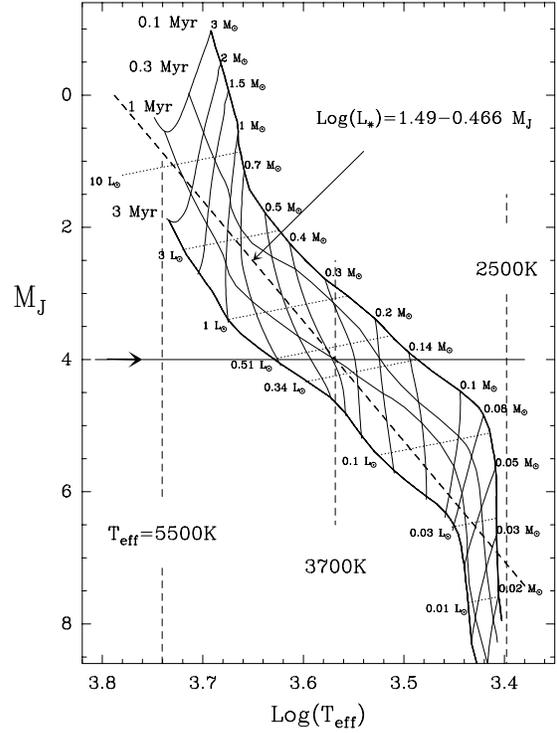}
\label{mjls}
\caption[]{
$M_J -$~log$_{10}$($\teff$) ``HR'' diagram showing the well-defined strip 
of the HR diagram occupied by the PMS tracks of D'Antona
\& Mazzitelli (1998). For 
$M_J = 4.0$ for instance, we see that log$_{10}$($\teff$) can span a range
of only $\sim 0.15$~dex for stellar ages between $0.1\,$Myr and 
$3\,$Myr.
We thus adopt the $M_J -$~log$_{10}$($\teff$) linear relationship displayed
as a heavy dashed line which corresponds to the conversion:
$\log_{10}(\ls) = 1.49-0.466\times M_{J}$.
}
\end{figure}

Furthermore, since  
bright $M_J$ sources tend to be of earlier spectral type than faint sources, 
we can in fact achieve more accurate luminosity estimates.  
Indeed, PMS stars are predicted to lie 
within a well-defined strip of the HR diagram. This 
is illustrated in Fig.~4 which displays 
model evolutionary tracks and isochrones from D'Antona \& Mazzitelli 
(\cite{dm97}) 
on a $M_J$--log$_{10}(\teff)$ diagram 
for PMS stars with ages between $0.1\,$Myr and $3\,$Myr. 
We have used the compilations of 
Hartigan et al. (\cite{hss94}), Kenyon \& Hartmann (\cite{kh95}), 
and Wilking et al. (\cite{wgm}) to derive an approximate linear 
interpolation for $BC_J$: 
$BC_J = 1.65+3.0\times$log$_{10}$($3700/\teff$). 
Figure~4 shows that, for a given observed value of $M_J$, the possible 
range of log$_{10}$($\teff$) is reduced to less than $\sim$~0.15~dex, inducing a 
maximum error of $\pm 23\,$\% ($\pm 0.09$~dex) on $\ls $. 
Based on Fig.~4, we have adopted a linear 
relationship between $M_J$ and $\teff$: 
log$(\teff/3700) = -0.055\times (M_{J}-4.0)$
(see the heavy dashed line in Fig.~4).
This leads to the following $\ls-M_J$ conversion:

\begin{equation}
\log_{10}(\ls) = 1.49-0.466\times M_{J}. \hspace{0.7cm}
\label{mj-lstar}
\end{equation}

The uncertainty 
resulting from this conversion should be on the order of 0.045~dex for 
$0.01 \le \ls \le 10.0\,\sol$. We therefore estimate that the total 
uncertainty on $\log_{10}(\ls)$  is 
$\sigma$({\small $\log_{10}\ls$}) 
$= (0.045^2+(0.466\times\sigma$({\small $M_J$})$)^2)^{1/2}$
$= 0.19$~dex.

The $\ls - M_{J}$ conversion described above cannot be applied 
to sources undetected in the J band. Instead, we use the H-band magnitude,
along with the extinction estimate derived from the $(H-K)$ color, but with
the additional complication that the circumstellar (disk) emission  
cannot be neglected. 
Meyer et al. (\cite{meyer97}) found that the H-band circumstellar excess 
is on the order of 20~\% of the stellar flux, on average, for the  
Taurus CTTS sample of Strom et al. (\cite{strom89})
(for a $\roph$ sample, see Greene \& Lada \cite{gl96}). 
This excess, expressed as a veiling index 
$r_{H} = F_{\nu}\!\!^H_{\hspace{0.1cm}\mathrm{exc}}
/F_{\nu}\!\!^{H}_{\hspace{0.1cm}\star}$ (e.g.
Greene \& Meyer \cite{gm95}), is equal to $\sim 0.2$. 
Accordingly, we have applied a systematic correction 
$\delta$H~$=\,-2.5\,\log_{10}(1+r_{H}) \sim 0.2$~mag.
The $\ls - M_{H}$ relationship obtained in a way similar to the 
J-band relation is then: 

\begin{equation}
\log_{10}(\ls)=1.26-0.477\times (M_{H}+0.2)
\end{equation}

The total uncertainty 
on $\log_{10}(\ls)$ derived from $M_H$ is then 
$\sigma$({\small $\log_{10}\ls$}) 
$= (0.045^2+(0.477\times\sigma$({\small $M_H$})$)^2)^{1/2}$
$= 0.29$~dex.

This method can also be applied to Class~III YSOs, using different values
for $(J-H)_0$ and $(H-K)_0$.
We have derived $\ls$
for all Class~III YSOs 
using the following relationships: $A_V  =  9.09 \times [(J-H)-0.6]$, or
$A_V  =  15.4 \times [(H-K)-0.15]$; and 
log$(\ls)  =  1.49-0.466 \times (J-0.265 \times A_V-5.73)$, or
log$(\ls)  =  1.26-0.477 \times (H-0.155 \times A_V-5.73)$ 
(the H-band IR excess for Class~III YSOs is negligible).\\
The resulting $M_J, M_H, A_V$, and $\ls$ estimates are listed in 
Table~3 for Class~II YSOs, and in Table~4 and Table~5 for Class~III YSOs.

\subsection{Disk luminosities for Class~II YSOs}

Since the SED of an embedded Class~II YSO peaks in the mid-IR 
range,
the ISOCAM fluxes should be approximately valid tracers of the total, 
bolometric luminosities ($\lbol$). 
To estimate $\lbol$ for weak Class~II sources, an empirical approach 
thus consists in using this $F_\nu^\mathrm{MIR}$--$\lbol $ relationship
after proper calibration on a sub-sample of (brighter) objects for which the 
luminosity can be derived by a more direct method. 
This approach has been adopted by, e.g., Olofsson et al. (\cite{ohk99}).
Here, we have used the $\ls$ estimates of Sect.~4.1 to check that 
a correlation is actually present between $\ls$ and the mid-IR 
fluxes. Figure~5 displays $\flwt$ (corrected for extinction) 
as a function of $\ls$ 
for the 104 Class~II sources detected both in the near-IR and in the mid-IR range. 

\begin{figure}[ht]
\centering
\includegraphics[width=8.5cm]{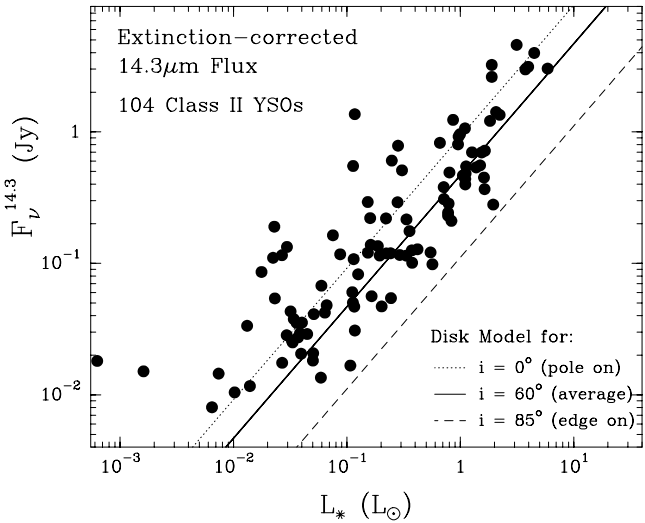}
\label{iso-ls}
\caption[]{
Correlation between $\flwt$ (corrected
for interstellar extinction with $A_{14.3}=0.03\times\av$)
and $\ls$. The observations are compared with a purely reprocessing 
disk model with $\ld=0.25\times\ls$ and a stellar 
contribution given by Eq.~6. 
The continuous line corresponds to an average disk inclination 
($i = 60\degr$), while the dotted and dashed lines correspond to 
more extreme inclinations ($i = 0\degr$ and $i = 85\degr$). 
}
\end{figure}

A correlation is found, showing that, despite some scatter, 
the ISOCAM fluxes can be used to give rough estimates of 
the stellar luminosities of Class~II YSOs. 
This is useful for the few ISOCAM sources of our
sample which have not been detected at near-IR wavelengths. 

The mid-IR emission of Class~II YSOs is usually interpreted as
arising from warm dust in an optically thick circumstellar disk. 
Using a simplified disk model (e.g. Beckwith et al. \cite{beckwith90}), 
it is easy to show that any observed monochromatic flux in 
the optically thick, power-law range of the disk SED
is simply proportional to the total disk luminosity $\ld$ 
divided by the projection factor cos($i$), where $i$ is the 
disk inclination angle to the line of sight. 
In the Beckwith et al. (1990) model, the disk is parameterized by a power-law 
temperature profile with three free parameters, $T_0$, $r_0$, and $q$,
such that: $T(r) = T_0 \times (r/r_0)^{-q}$. 
Here, we have adopted $T_0 = 1500$~K, meaning that the disk inner radius 
is at the dust sublimation temperature, and $q=2/3$, corresponding 
to an IR spectral index $\air=2/q-4=-1.0$ typical of CTTS spectra.

We must, however, account for the fact that the stellar emission
itself is not completely negligible
in the mid-IR bands, especially at 6.7~$\mu$m.
A simple blackbody emission at $T_\star=3700$~K
gives 
\begin{eqnarray}
F_{\nu\hspace{0.1cm}\star}^{\,\,\,6.7}=0.082\times(\ls/1\,\sol)\,\,\mathrm{Jy} \\ 
\flwts=0.021\times(\ls/1\,\sol)\,\,\mathrm{Jy}.
\end{eqnarray}

\noindent
Since $\flwts $ is significantly smaller than $\flwds $, 
only $\flwt$ is used to estimate $\ld $.
Assuming cos$(i)=0.5$ ($i=60\degr$) in all cases, we have computed 
$\ld$ for each of the Class~II YSOs with $\ls$ and $\av$ estimates 
from Sect.~4.1, based on the following relationship:   
 
\begin{equation}
\ld/1\,\sol  =  (\flwt\times
10^{+0.4\times A_{14.3}}-\flwts)/1.80\,\mathrm{Jy}, 
\end{equation}

\noindent
where $A_{14.3}\!=\!0.03\times\av$ and 
$\flwts$ from Eq.~6. The results are given in the
last column of Table~3.

According to this model, the mid-IR flux is a direct tracer of the 
disk luminosity, and the $F_\nu^\mathrm{MIR} - \ls$ correlation
of Fig.~5 simply expresses that $\ld$ correlates with $\ls$.
The origin of $\ld$ is either the release of gravitational 
energy by accretion in the disk, 
or the absorption/reprocessing of stellar photons by the dusty disk. 
In the latter case, 
$\ld$ is naturally proportional to $\ls$.
The fraction $\ld/\ls$ of stellar luminosity reprocessed by the disk
depends on the spatial distribution of dust. 
In the ideal case of an infinite, spatially flat
disk, this fraction is 0.25 (Adams \& Shu \cite{as86}). If the disk is 
flared, 
$\ld/\ls$ is larger, while it is smaller if the disk has a inner hole. 
The theoretical $F_{\nu}^\mathrm{MIR}-\ls$ correlations 
plotted in Fig.~5 correspond to $\ld=0.25\times\ls$ and to three 
representative inclination angles. 
The fact that this simple 
model accounts for the 
observed correlation quite well, 
suggests that the disks of most $\roph$ Class~II YSOs are passive disks
dominated by reprocessing. 
This is consistent with recent estimates of the disk accretion level
in Taurus CTTSs (e.g. Gullbring et al. \cite{gullbring}).
The typical disk accretion rate of a CTTS is estimated to be 
$10^{-8}\,\sm/$yr, corresponding to an accretion luminosity 
$\lacc=0.025\,\sol$ for $\rs=3\,\sr$ and $\ms = 0.25\,\sm$ (i.e., 
$\ls \sim 0.25\,\sol $ at 1$\,$Myr). 
In this case, the luminosity due to reprocessing is $\sim 5$ times 
larger than the accretion luminosity in the disk.

On the other hand, 37 sources (among a total of 104) are located 
above the passive disk model lines in Fig.~5.
These are good candidates for having an active disk with an accretion 
rate typically larger than 
$\sim 5\times(\ls/0.25\sol)\times10^{-8}\,\sm/$yr. 

Overall, we find that the median $\ld/\ls$ ratio is 0.41 for the 
93 Class~II sources detected in the near-IR and 
with $\flwt>\,15\,$mJy (i.e. the completeness level derived in
Sect.~2.4). 
Using this ratio, we have derived rough estimates of the
stellar luminosities of the 15 Class~II YSOs which have no 
near-IR photometry (see Table~3) as follows:

\begin{equation}
\ls(\sol)\approx \ld/0.41\approx (0.97\times1.6/0.41)\times
(\flwt/1.8\,\mathrm{Jy}).
\end{equation}

[The factor 0.97 corresponds to a typical stellar contribution 
of $3\,$\% at 14.3~$\mu$m, while the factor 1.6 is an average 
extinction correction at 14.3~$\mu $m (corresponding to 
$<A_V> = 17$~mag).]

\subsection{Calorimetric luminosities for Class~I sources}

The most direct method of estimating the total 
luminosities $\lbol$ of embedded YSOs consists in integrating the 
observed SEDs (cf. WLY89).
However, since most of the $\rho$~Ophiuchi Class~II and Class~III YSOs 
are deeply embedded within the cloud ($A_V \simgt 10$), only
a negligible fraction of their bolometric luminosity can be recovered
by finite-beam IR observations (e.g. Comer\'on et al. \cite{crbr}).
We thus do not attempt to derive calorimetric estimates of $\lbol$ 
for these sources.
In contrast, the calorimetric method is believed to be appropriate 
for Class~I YSOs since these are self-embedded in substantial 
amounts of circumstellar material which re-radiate locally
the absorbed luminosity (cf. WLY89 and AM94).
Using our new mid-IR measurements, we have evaluated 
the calorimetric luminosities ($\lcal$) of the 16 Class~I YSOs 
observed in our survey. 
Only 7 of them have reliable IRAS fluxes
up to 60 or 100~$\mu$m (IRS54, IRS44, GSS30, IRS43, EL29, IRS48,
IRS51). 
For these, the median of the ratio of 
$\lcal $(6.7--14.3$\mu$m)
to $\lbol$
is found to be 9.8, suggesting that the typical fraction of a  
Class~I source's luminosity radiated between 6.7
and 14.3~$\mu$m is $\sim 10$~\% .
Assuming that this ratio is representative of all Class~I YSOs, 
we have derived estimates of $\lbol$ for the 
remaining 9 weaker Class~I sources (i.e., CRBR85, LFAM26, LFAM1, WL12, IRS46, 
CRBR12, IRS67, CRBR42, WL6). These luminosities are listed in Table~2.

\subsection{Luminosity function of the $\roph$ embedded cluster}

Combining the $\ls $ luminosities determined in Sect. ~4.1 for the 
sources detected in the near-IR with the $\ls$ estimates from $\ld$ 
for the sources without near-IR measurements (Sect.~4.2), 
we have built a luminosity function for Class~II YSOs which represents 
a major improvement over previous studies (see Fig.~6a).
 In terms of $\ls$, the completeness level for this population    
can be estimated from the $\flwt$ completeness limit 
derived in Sect.~2.4 ($\flwt \sim 15\,$mJy) 
using Eq.~8: $\ls^{comp} (\rm {Class II}) = 0.032\,\sol$. 
While the luminosity function previously published by 
Greene et al. (\cite{gwayl}) included only 33 (bright) Class~II sources
and suffered from severe incompleteness below  $\ls \sim 1-2\,\sol$, 
our present completeness level 
is a factor $\sim $~30--50 lower.
The new luminosity function shows a marked flattening 
in logarithmic units at $\ls \sim 2\,\sol$, 
well above our completeness limit.
This important new feature is discussed in Sect.~5 below.
 
\begin{figure*}[ht]
\hspace{1cm}\includegraphics[width=8.9cm]{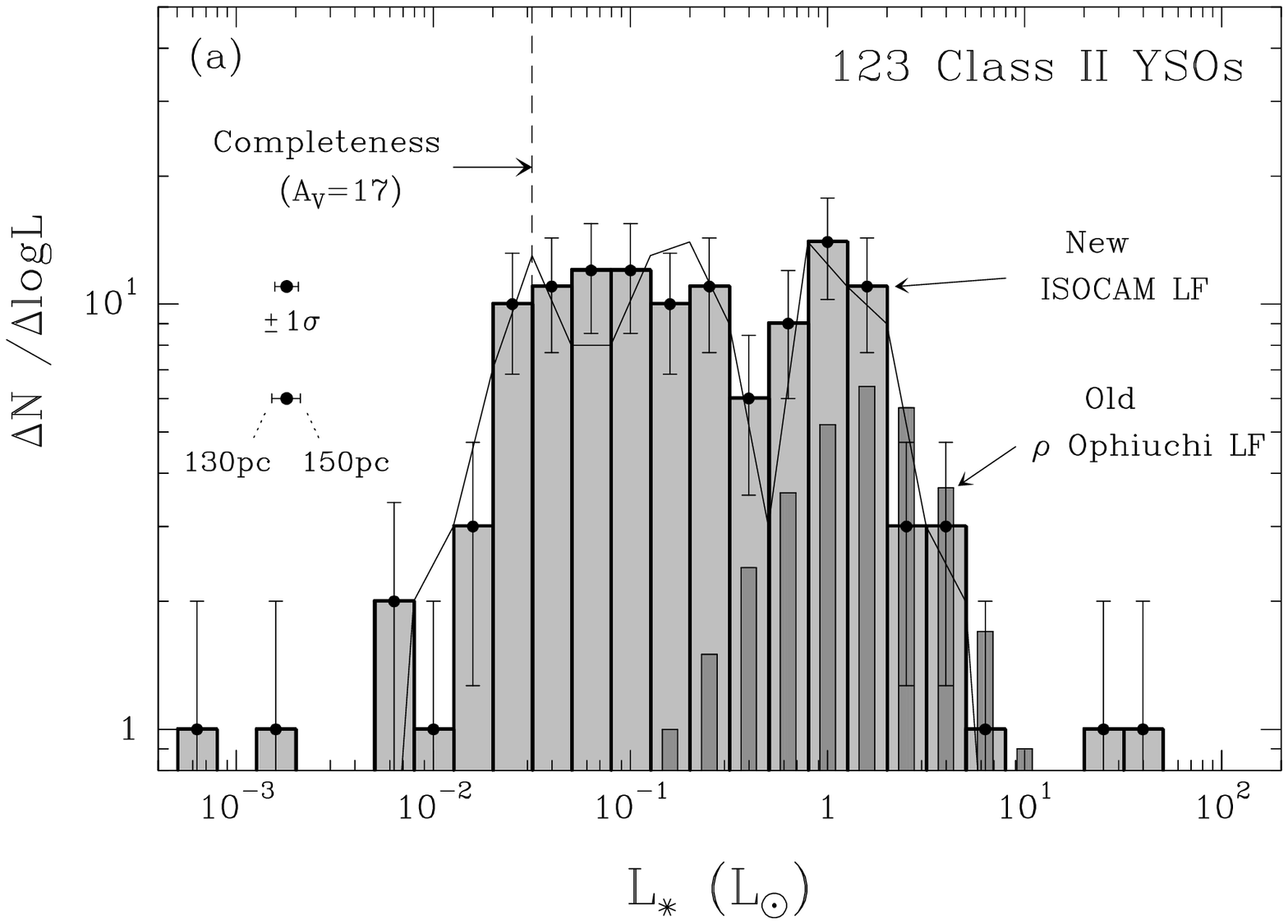}
\hspace{0.5cm}\includegraphics[width=5.8cm]{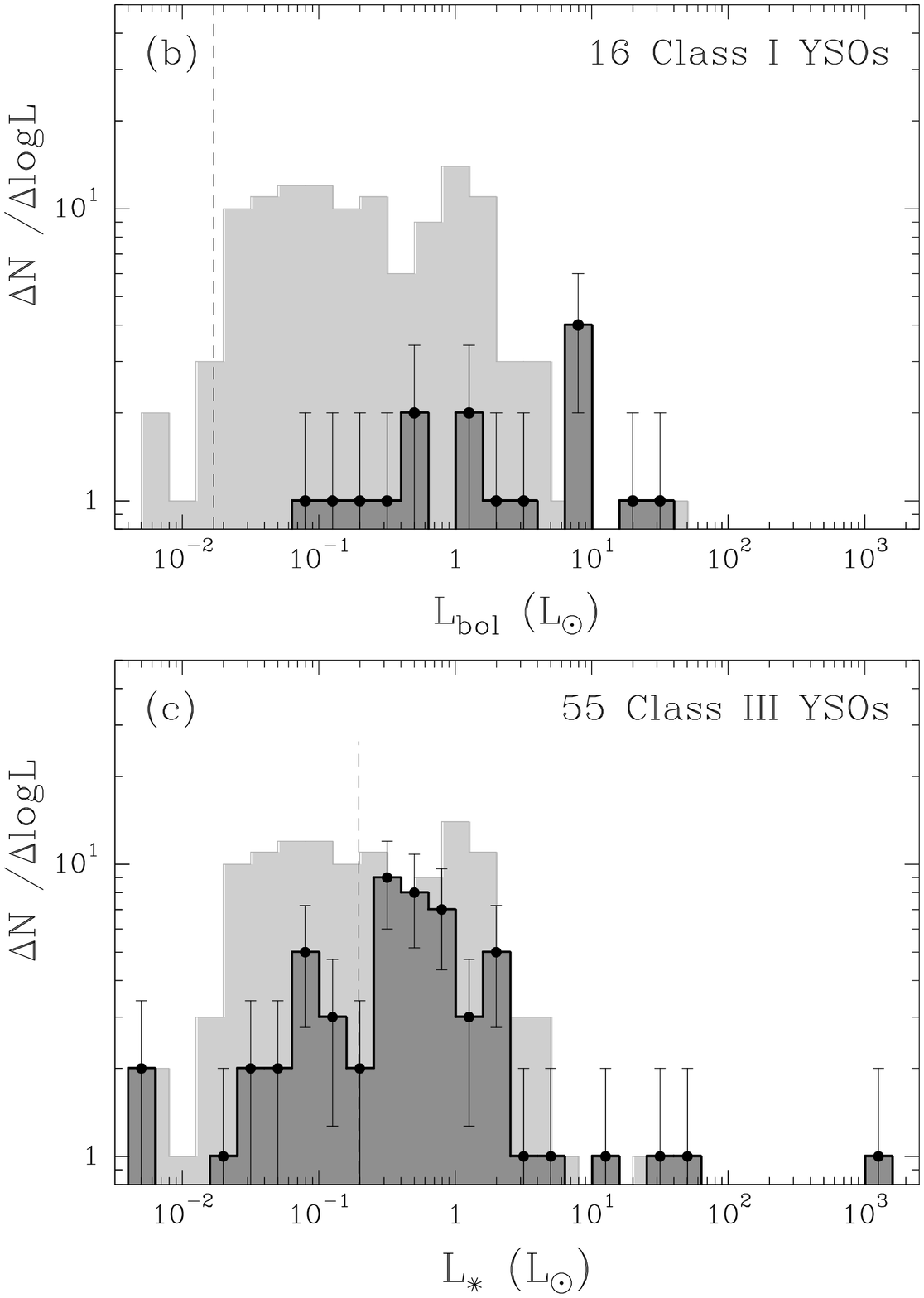}
\label{lumf}
\caption[]{
Luminosity functions (LF) {\bf (a)} for the 123 Class~II YSOs 
(continuous histogram with statistical 
error bars). 
The function corresponding to a similar histogram
shifted by half the 0.2~dex bin size  
is shown as a thin curve to illustrate the level of 
statistical fluctuations due to binning. 
The LF of 33 Class~II sources from Greene et al. (1994) 
is displayed as a darker histogram 
(rebinned to 0.2 dex bins, and rescaled to $d=140$~pc 
for better comparison with the new LF). 
The typical $\pm 1\sigma$ uncertainty on the position of each 
bin resulting from the individual uncertainties on $\ls$ 
is indicated, along with the effect of the distance uncertainty;
{\bf (b)} for the 16 Class~I YSOs ; 
and {\bf (c)} for the sample of 55 (19 confirmed $+$ 36 candidate) 
Class~III sources 
from Table~4 and 5.
The LF of (a) is shown in the background for comparison.
The dashed vertical lines show the respective completeness levels.
}
\end{figure*}

Based on the $\lcal $ estimates of Sect.~4.3, 
a new bolometric luminosity function for the 
16 Class~I YSOs of $\roph$ is displayed in Fig.~6b. 
The associated completeness level is 
derived from $\flwd = 10\,$mJy and $\flwt = 15\,$mJy (Sect.~2.4) 
using $\lbol /\lcal $(6.7--14.3$\mu$m)$\approx 9.8$ (Sect.~4.3):
$\ls^\mathrm{comp} (\rm {Class I}) = 0.017\,\sol$.
The median $\lbol$ for Class~I YSOs 
is $1.6\ \sol$, which is $\sim 8$ times larger 
than the median $\ls$ of Class~II YSOs ($0.20\,\sol$). 
The luminosities of Class~I YSOs span a range of two orders of magnitude 
between $\sim 0.1\ \sol$ and $\sim 10\ \sol$, which is roughly as wide as 
the luminosity range spanned by Class~II YSOs. 
The comparatively large value of $\lbol$ for Class~I YSOs is probably
due to a dominant contribution of accretion luminosity 
as expected in the case of protostars.

In Fig.~6c, we plot the luminosity function of 
the 55 Class~III sources that are located within the CS contours 
of Fig.~1 and for which we have enough near-IR data 
to derive $\ls$ according to the procedure described in Sect.~4.1.
This sample comprises 19 confirmed Class~IIIs from Table~4 together 
with 36 candidate Class~III sources from Table~5.
It might be contaminated by a few background/foreground sources but 
is characterized by a relatively well defined completeness luminosity. 
From the completeness level $\flwd = 10\,$mJy 
and using Eq.~5 with an average extinction
correction corresponding to $A_V = 17$~mag, we get
$\ls^{comp} (\rm {Class III}) = 0.20\,\sol$, 
which is $\sim 6$ times higher than $\ls^{comp} (\rm {Class II})$.
Deep X-ray observations with $XMM$ 
should improve the completeness luminosity for Class~III YSOs 
by an order of 
magnitude in the near future (cf. discussion by Grosso et al. \cite{grosso} 
in a companion paper).

\section{Modeling of the luminosity function}

If the mass-luminosity relationship of PMS stars were a simple,  
fixed power-law function, the shape of the luminosity function 
would directly reflect the underlying mass function. 
Unfortunately, the mass-luminosity relation  
is a complex function with inflexion points and is strongly
age-dependent. Deriving the mass function from the luminosity function 
thus requires knowledge of the stellar age distribution.
The $\roph$ embedded cluster is believed to be  
younger than most known star clusters (e.g. Luhman \& Rieke \cite{lr99}), 
with a typical age on the order of, or less than, 1~Myr 
(e.g. WLY89, Greene \& Meyer \cite{gm95}). 

The 123 Class~II sources identified with ISOCAM are presently  
the most complete sample of young stars available in 
the cluster.
The corresponding mass function is estimated down to $0.055\,\sm$ 
in Sect.~5.1.  
We investigate 
the effect of including Class~III YSOs by modeling the luminosity function of 
the 135 Class~II and Class~III objects located inside the CS contours 
of Fig.~1 (Sect.~5.3).

\begin{figure*}[ht]
\centering
\includegraphics[width=13cm]{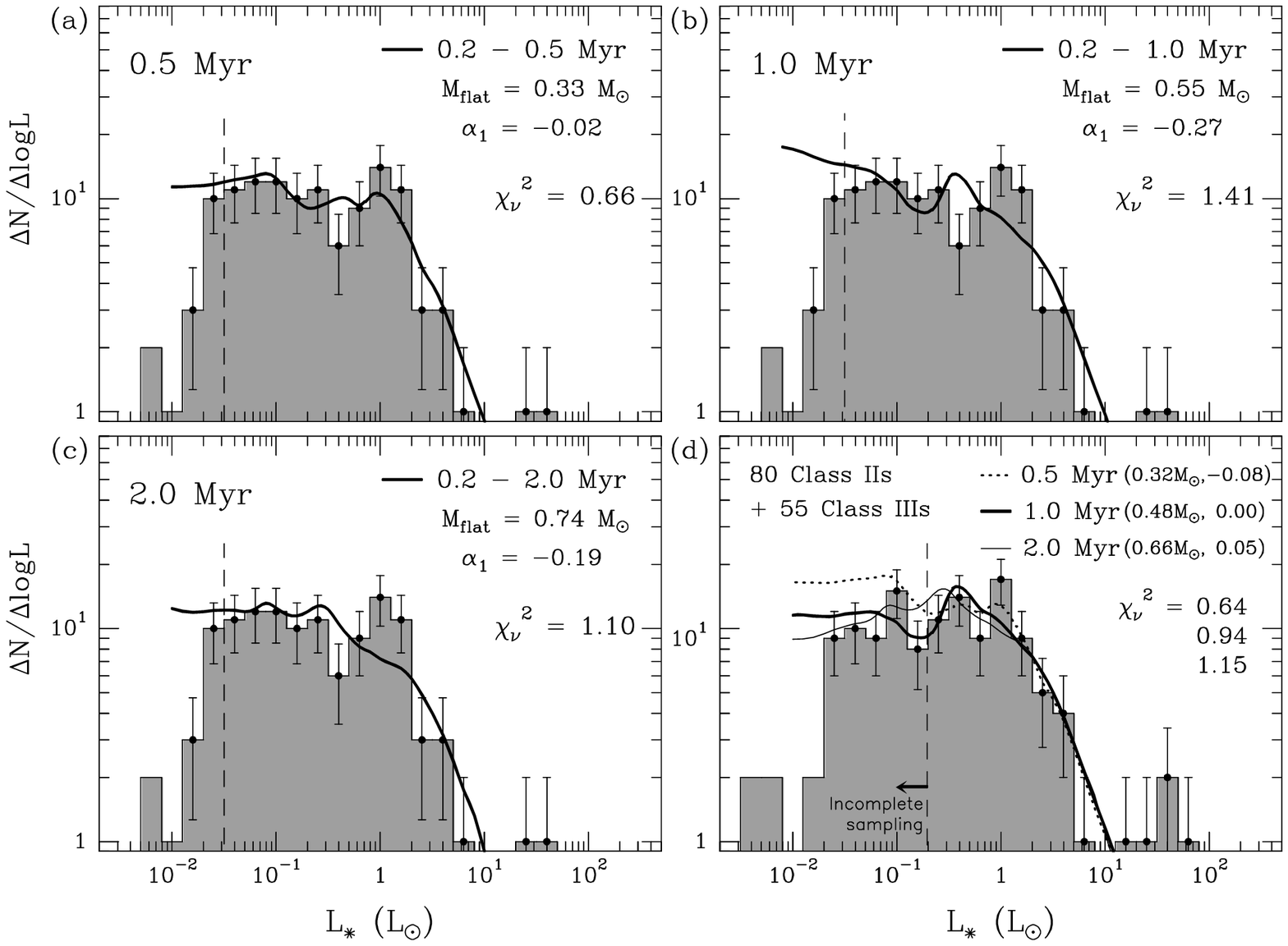}
\caption[]{
Best model fits (heavy curves) to the luminosity function of 
Class~II sources obtained for a stellar age distribution corresponding to 
a constant formation rate over {\bf (a)} 0.5 Myr,
{\bf (b)} 1 Myr, and {\bf (c)} 2 Myr (the first
0.2~Myr have been avoided in the PMS tracks to account for a 
typical protostellar timescale).
In {\bf (d)}, similar best fits are shown for the combined luminosity 
function of 135 Class~II and Class~III sources inside the
CS contours of Fig.~1.
The vertical dashed lines show the completeness levels
of the respective luminosity functions (see Fig.~6).
}
\label{lfmod}
\end{figure*}

\subsection{Constraints on the mass function of Class~II YSOs}

We have modeled the observed luminosity function using a two-segment 
power-law mass function with two free parameters, $\mflat$ and $\alpha_1$:
{\small d$N/$d$\,$log$\ms  \propto \ms^{\alpha_1}$} for $\ms<\mflat$, 
and {\small d$N/$d$\,$log$\ms \propto \ms^{-1.7}$} for
$\ms>\mflat$. The high-mass range (for
$\ms>\mflat$) is simply taken to be a power-law with index 
$-1.7$ (whereas the Salpeter index would be $-1.35$), in agreement with 
the IMFs favored by Kroupa et al. (\cite{kroupa}) and Scalo (\cite{scalo}). 
We assume that the cluster forms stars with a fixed IMF, independent of 
time. The evolutionary tracks of D'Antona \& Mazzitelli 
(\cite{dm97})\footnote{We here adopt the tracks of D'Antona \& 
Mazzitelli (\cite{dm97}) as they provide the most  
extensive set of PMS models available to date, including a tight grid 
of masses and ages.
In the future, we plan to check the details of our results using other 
tracks such as those of Baraffe et al. (\cite{b98}) or 
Palla \& Stahler (\cite{ps99}).} 
are used to derive a set of mass-luminosity relations adapted to the $\roph$
PMS population, under the following three hypotheses about the star
formation history: 
Constant star formation rate over a) 0.5~Myr (Fig.~7a), 
b) 1~Myr (Fig.~7b), and c) 2~Myr (Fig.~7c). 
Cases b) and c) correspond to the simplest scenarios consistent with 
current observational constraints (e.g. WLY89). Case a) mimics a recent
``burst'' of star formation suggested by recent
studies (e.g. Greene \& Meyer \cite{gm95} and Luhman \& Rieke \cite{lr99}).  
These three scenarios are roughly representative of our current (imperfect)
knowledge of the stellar age distribution in the cluster.

The fitting analysis is performed on the 12 (logarithmic) luminosity bins 
above the completeness level of 0.032$\,\sol $ and below $10\,\sol$ 
by varying $\alpha_1$ and $\mflat$. 
Due to the age ambiguity for a 
particular ($\ms$, $\ls$) in the PMS tracks for $\ms\simgt 2\,\sm$ and 
$age < 3\,$Myr as a result of
the transition from convective to radiative interiors (e.g. 
Fig.~4), our simplified modeling is not valid for $\ls\simgt10\,\sol$. 
The best fit is shown as a heavy curve
for each of the three assumed age distributions (Fig.~7a, 7b, and 7c).
The flattening of the luminosity function below $2\ \sol $ is well 
reproduced in all three cases.  
The best-fit model in the 1~Myr case (Fig.~7b) is however not quite as 
good as the two others since it predicts a peak at $\ls \sim 0.4-0.5\,\sol$ 
where the data show a dip. This is reflected
by the largest value $1.41$ for the reduced 
$\rchi \equiv \frac{\chi^2}{N-n}$ (with $N=12$ fitted data points and 
$n=2$ free parameters).
The model of Fig.~7c reproduces the data somewhat better 
($\rchi = 1.10$) as the predicted peak luminosity 
moves down to $\ls \sim 0.2-0.3\,\sol$.
A marginally better fit is obtained with the 0.5~Myr model  
(Fig.~7a, $\rchi = 0.66$).
In this case, the model luminosity function has a local maximum  
at $\ls\simgt1\,\sol$ as observed and the flattening for $\ls \sim 2\,\sol $
is particulary well reproduced.

The best-fit values of $\mflat$ along with the formal $1\,\sigma$ 
errors resulting from the fitting procedure are as follows:
$0.33\pm0.04\,\sm$, $0.55\pm0.09\,\sm$, and $0.74\pm0.13\,\sm$ 
for Fig.~7a, 7b, and 7c, respectively.
These $\mflat$ values follow the expected trend that the older the stars, 
the higher the derived masses for the same luminosities.  
The final error on the determination of $\mflat$ is clearly dominated by the 
uncertainties on the stellar age distribution.
We adopt the following average value: $\mflat = 0.55 \pm 0.25\,\sm$.
The best-fit values for $\alpha_1$ are $-0.02\pm0.15$, $-0.27\pm0.13$, 
and $-0.19\pm0.13$, respectively. There is no correlation with age in this
case.
We conservatively adopt $\alpha_1 = -0.15 \pm 0.2$.
These constraints on the mass function are valid for 
$0.032\,\sol < \ls < 10\,\sol$, which approximately corresponds to
$0.055\,\sm \simlt \ms \simlt 2\,\sm$.

Under the same three assumptions about the stellar age distribution as 
above, we have also derived a range of stellar masses directly from the 
$M_J$ magnitude for each of the 123 Class~II YSOs. We display the results
as a mass function in Fig.~8, where the vertical error bars reflect the 
uncertainties induced by the various age assumptions rather than the 
statistical errors.
The best two-component power-law mass function derived above 
is superposed as a heavy curve.
The error bars are particulary large for the two mass bins close to 1$\,\sm$, 
i.e., just above $\mflat$.
This is due to a shift of a large number of stars beyond $\mflat$ 
when the stellar ages are varied upward.
The global shape of the mass function is however fairly
well determined: It is basically flat (in logarithmic units) 
at low masses down to 0.055$\,\sm$ and shows a steep decline beyond 
$\sim 0.5\,\sm$.
 
\begin{figure}[ht]
\centering
\includegraphics[width=8.0cm]{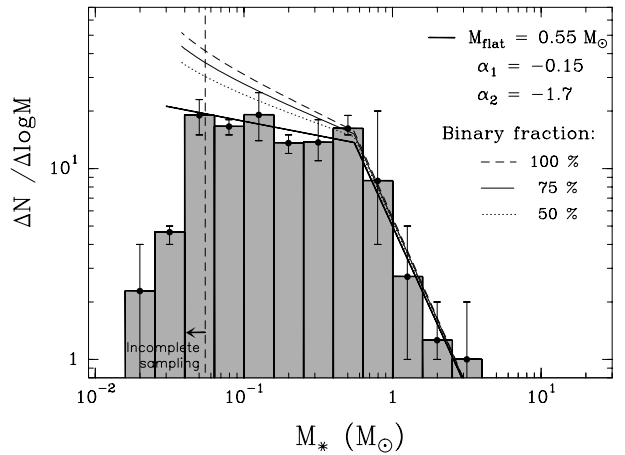}
\caption[]{Mass function of the 123 Class~II YSOs. 
The range of possible age distributions 
(see Sect.~5.1) induces a range of masses for each star and thus
an uncertainty on the derived histogram displayed as vertical
error bars. 
The best two-segment power-law mass function of 
Sect.~5.1 (heavy solid curve), and the effect of 
binarity as a function of the binary fraction from 50 \% to 100~\% (light 
curves -- see Sect.~6.2) are shown.  
The vertical dashed line marks the completeness level of $0.055\,\sm$.
}
\label{binary}
\end{figure}

\subsection{Possible deuterium feature in the luminosity function of Class~II YSOs}

Some of the features (peaks and dips) apparent in the observed 
luminosity function might be real as they are also predicted by the models
(see Fig.~7).
The feature expected in the luminosity function
of young stars as a result of 
deuterium burning during PMS contraction is particularly interesting
since its location is age-dependent 
(e.g. Zinnecker et al. \cite{zinnecker}).
As already noted, the model fit obtained under the 0.5~Myr scenario 
is particularly good 
as it predicts a peak at $\ls\sim 1.5\,\sol$, consistent with the 
observations (see Fig.~7a).
In the model, this $1.5\,\sol$ peak corresponds to an inflexion point in 
the mass-luminosity relation due to deuterium 
burning in $\sim 0.4$-Myr old stars with $\ms \sim 0.5\,\sm$.
If this peak in the luminosity function is confirmed,
it will provide strong evidence that the population of Class~II YSOs
is particularly young in $\roph$. 

\subsection{Effect of Class~III YSOs}

There is most probably a significant population of Class~III objects 
embedded in the cloud (e.g. Sect.~3.5). To evaluate how the 
presence of this population may affect our conclusions 
on the IMF of the cluster (see Sect.~6 below), we
here consider the luminosity function of a combined sample of Class~II
and Class~III sources. The sample comprises the 80 Class~II and 55 Class~III
YSOs located inside the CS contours of Fig.~1. It should be complete 
down to $\sim 0.2\,\sol$. (Note, however, 
that only 19 of the 55 Class~III sources of this sample
are confirmed YSOs at this stage -- see Sect.~4.4.)
The associated luminosity function, displayed in Fig.~7d, 
is very similar to the luminosity function of Class~II YSOs. 
In particular, it exhibits a flattening below $\sim 1-2\,\sol$. 

The results of a modeling similar to that of Sect.~5.1, but performed
on only 8 luminosity bins between 0.2 and 10$\,\sol$, are
shown in Fig.~7d. The best fits found under the same three star 
formation scenarios as in Sect.~5.1 are good (with $\rchi = 0.64$, 0.94, 
and 1.15, respectively).
The best-fit values for the parameters $\mflat$ and $\alpha_1$ 
of the mass function are as follows:
$\mflat=0.32\pm0.04\,\sm$, $0.48\pm0.08\,\sm$,
$0.66\pm0.10\,\sm$, and $\alpha_1=-0.08\pm0.45$, $0.00\pm0.35$, 
$+0.05\pm0.19$ for star formation durations of 0.5, 1, and 2~Myr, 
respectively. Combining these results, we obtain  
$\mflat \sim 0.5\pm0.2\,\sm$ and $\alpha_1\sim0.0\pm0.4$. 
Within the uncertainties these values are essentially identical to those found 
in Sect.~5.1 for the sample of Class~II YSOs, although they apply to a more
limited mass range, approximately $0.17\,\sm \simlt \ms \simlt 2\,\sm$.

In contrast to the Class~II case, the luminosity function of the 
Class~II$-$Class~III combined sample does not show a peak at 
$\ls = 1.5\,\sol$ (cf. Fig.~7d). 
This may be understood if the average stellar age of the combined population
is somewhat larger than that of the population of Class~II sources 
(as is expected).

\section{Discussion}

\subsection{The initial mass function in $\roph$}

Since the $\roph$ molecular cloud is still actively forming stars, 
the final IMF of the cluster cannot be directly measured. 
The masses of the stars already formed 
in the cluster 
provide only a ``snapshot'' of the local IMF (cf. Meyer et al. 
\cite{meyer00}).
However, assuming that the mass distribution of formed stars does not 
change significantly with time during the cluster's history 
(an assumption made in the models of Sect.~5.1), any 
snapshot of the mass distribution taken on a large, complete population of
PMS objects should accurately reflect the end-product IMF. 

Compared to previous investigations of the IMF 
in the $\roph$ cloud based exclusively on 
near-IR data (e.g. Comer\'on et al. \cite{crbr}, 
Strom et al. \cite{sks}, Williams et al. \cite{williams}, 
Luhman \& Rieke \cite{lr99}), 
the use of mid-IR photometry 
has allowed us to consider a much larger sample of young stars 
(see, e.g., Fig.~6).
Since our sub-sample of Class~II YSOs is fairly large (123 objects) 
and complete 
down to low luminosities, it provides an excellent opportunity to derive
improved constraints on the $\roph$ IMF down to low masses.
 
In a statistical sense at least, 
Class~II YSOs are believed to represent a specific 
phase of PMS evolution which follows the (Class~0 and Class~I) 
protostellar phases, and precedes the Class~III phase (see 
Sect.~1).
Due to their short lifetime ($\simlt 10^5\,$yr), protostars make up 
only a small fraction of a young cluster's population 
(cf. Fletcher \& Stahler 1994), and can be neglected in the global 
mass function. Furthermore, in contrast to PMS stars, protostars
have not yet reached their final stellar masses.

Class~III objects are more numerous and should thus contribute significantly 
to the global mass distribution. Furthermore, it has been suggested that 
some YSOs evolve quickly to the Class III stage, perhaps as early as the
``birthline'' for PMS stars (e.g. Stahler \& Walter 1993), 
and spend virtually no time in the Class~II phase 
(cf. Andr\'e et al. 1992, Greene \& Meyer 1995). Since such objects cannot be 
identified through IR observations, their exact number and mass distribution
will not be known until the results of deep X-ray (and follow-up) surveys are
available (see Sect.~3.5).
However, providing the (range of) evolutionary timescale(s)  
from Class~II to Class~III is independent of mass, both 
classes of PMS objects should have identical mass functions. 
The results of Sect.~5.3 do seem to support this view, as they suggest  
that the mass functions of the Class~II and Class~III samples 
do not differ down to $\ms \sim 0.17\,\sm$.

We therefore conclude (and will assume in the following) 
that the mass distribution of Class~II YSOs 
determined in Sect.~5.1 and shown in Fig.~8 currently represents our best 
estimate of the IMF in the $\roph$ embedded cluster down to 
$\ms \sim 0.055\,\sm$. As discussed in Sect.~6.2 below, this mass 
function applies to stellar systems rather than single stars.

\subsection{Influence of binary stars}

A large proportion ($\simgt 50\% $) of field stars are in fact multiple 
(e.g. binary) systems (e.g. Duquennoy \& Mayor \cite{dm91}).
This is also true for young PMS stars, and there is a growing body 
of evidence that the binary fraction is even larger in a young PMS cluster 
like $\roph$ than for main sequence stars in the field 
(e.g. Leinert et al. \cite{leinert93}; Simon et al. \cite{simon95}).
Since the present study is based on ISOCAM/near-IR observations which 
do not have enough angular resolution to separate most of the expected 
binaries, a significant population of low-mass stellar companions 
are presumably missing from the mass function derived above. 
These low-mass companions are hidden by the corresponding primaries. 

To estimate the magnitude of this binary effect, we show three simple 
models in Fig.~8 which assume a population of hidden secondaries 
corresponding to a 
binary fraction $f$ of 50\%, 75\%, and 100~\%, respectively. 
In each case, we start from a primary mass function 
with the two-segment power law form 
derived in Sect.~5.1. We then add a population
of secondaries with 
masses distributed uniformly (in logarithmic units) between a minimum mass 
of $0.02\,\sm$ and the mass of the primary. 
We thus assume that the 
component masses are uncorrelated and drawn from essentially the same
mass function (cf. Kroupa et al. \cite{kroupa}). 

The global mass functions resulting from addition of companions to 
the primary mass function are shown in Fig.~8.
It can be seen that the global mass functions are similar
in form to the original primary mass function. 
Neither the position of the break point ($\mflat$) nor the slope in the
high-mass range are affected by the addition of companions. 
However, the slope in the low-mass range ($\alpha_1$) steepens as the
binary fraction increases. Indeed, the power-law index between $0.055\,\sm$ 
and $\mflat=0.55\,\sm$ changes from $\alpha_1 = -0.15$ to  
$\alpha_1 = -0.31$ for $f=50\% $, $\alpha_1 = -0.37$ for $f=75\% $, 
and $\alpha_1 = -0.42$ for $f=100\% $. 

In summary, if we account for uncertainties in 
the binary fraction ($50\% \simlt f \simlt 100\% $), our 
best estimate of the single-star mass function in $\roph$ is
well described by a two-segment power-law with a 
low-mass index $\alpha=-0.35\pm0.25$ down to $0.055\,\sm$, 
a high-mass index $\alpha_2 = -1.7$, and a
break (flattening) occurring at $\mflat \sim 0.55\pm0.25\,\sm$. 

\begin{figure}[ht]
\centering
\includegraphics[width=8.8cm]{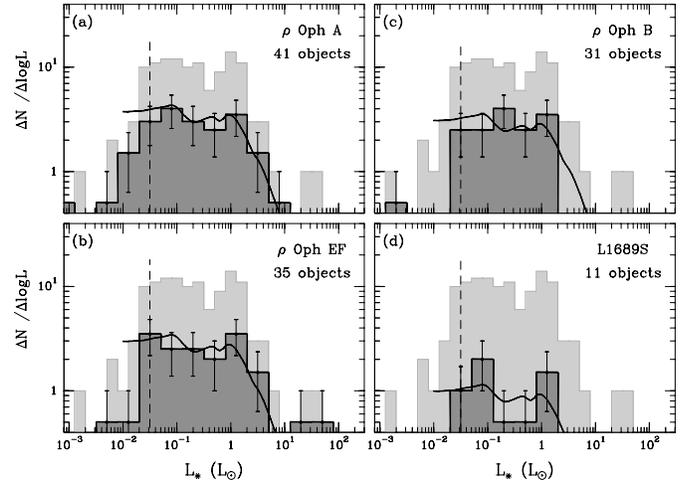}
\caption[]{
Individual luminosity functions for the four sub-clusters associated with the
dense cores Oph~A, Oph~B, Oph~EF, and L1689S (see Fig.~1).
These are displayed with 0.4~dex bins and scaled by a factor 1/2 for direct 
comparison with the 0.2~dex bin total luminosity function  
(light grey histogram in the background). The light solid curves correspond to 
the model of Fig.~7a scaled to the number of stars in each sub-cluster.
}
\label{cluster}
\end{figure}

\subsection{Luminosity/mass functions in individual cloud cores}

Our sample of Class~II YSOs is large enough 
that we can study the properties of the four sub-clusters 
associated with the dense cores Oph~A, Oph~B, Oph~EF, 
and L1689S (see Fig.~1). All 123 Class~II sources but 5 belong to 
these 4 sub-clusters.
The luminosity functions of Class~II objects in the individual sub-clusters 
are displayed in Fig.~9 along with the `best-fit' model of Sect.~5.1 
(solid curve).
They all agree reasonably well in shape with both the total luminosity 
function and the model: all four luminosity functions are essentially flat 
over two orders of magnitude in luminosity and appear to have a peak at 
$\ls \sim 1.5\,\sol$. 
The agreement is particularly good for the sub-cluster with the largest 
number of stars, Oph~A (see Fig.~9a), but even the smallest sub-cluster,  
L1689S, tends to reproduce the shape of the global luminosity function 
on a smaller scale (Fig.~9d). 
In contrast, for instance, 
the luminosity function derived for the Chamaeleon I cloud based on
ISOCAM data (Persi et al. \cite{pmo00}) differs markedly from 
the $\roph$ luminosity functions. It does not exhibit any peak 
at 1.5$\,\sol$ and is consistent with an older ($\sim 3$~Myr) 
PMS population having a similar underlying mass function (see
Kaas \& Bontemps \cite{kb01}).

The similarity of the individual luminosity functions suggests similar
distributions of stellar ages and stellar masses in each of the 
four sub-clusters. 
 
\subsection{Global aspects of star formation in $\roph$}

After correction for unresolved binaries (assuming a binary 
fraction $f=75\% $), the total number of Class~II sources (including 
companions) down to 0.055$\,\sm$ is $\sim $~145. 
Assuming a Class~III to Class~II number ratio of 19/22 as found by X-ray 
surveys (Grosso et al. \cite{grosso}
-- see Sect.~3.5), we infer the presence of $\sim 125$ Class~III stars 
(including associated companions) in the 
same mass range.
The typical number ratio of Class~Is (plus Class~0s) to Class~IIs is 
18/123 suggesting an additional $\sim 21$ embedded YSOs. 
Altogether, we therefore estimate that there are currently 
$N_\star \sim 291$ YSOs
down to $\sim 0.055\,\sm$ including $\sim 19$~\% of brown dwarfs.
The average and median masses of these objects are $\sim 0.35\,\sm$
and $\sim 0.20\,\sm$ respectively.
The total mass of condensed objects (including brown dwarfs) in 
the cluster is thus estimated to be 
$\ms^\mathrm{clust}=291\times0.35\sim102\,\sm$. 
(The brown dwarfs with $0.055\,\sm < \ms < 0.08\,\sm$ contribute only 
$\sim 4\,\%$ of this mass.)
Restricting ourselves to L1688 (thus subtracting the 
$\sim 10\,\%$ contribution from L1689) whose average radius is approximately 
0.4~pc (cf. CS contours in Fig.~1), we find 
$N_\star^\mathrm{L1688} \sim 262$, 
$\ms^\mathrm{L1688} \sim 92\,\sm$, 
$n_\star^\mathrm{L1688} \sim 980$~stars/pc$^3$, and 
$\rho_\star^\mathrm{L1688} \sim 340\,\sm$/pc$^3$, where $n_\star$ and 
$\rho_\star$ are the stellar number density and stellar mass (volume) density 
of the cluster, respectively.

Adopting a conservative value of 2~Myr for the cluster age, the total 
mass of $\ms^\mathrm{clust}=102\,\sm$ translates into  
an average star formation rate of $5.1\times10^{-5}\,\sm/$yr, corresponding to
one new YSO (of typical mass 0.20$\,\sm$) every $\sim 4000$~yr.
 
Lastly, we can derive the star formation efficiency (SFE) 
in L1688, defined as SFE$=\mst/(\mst+\mgas)$. The total molecular gas 
mass, $\mgas$, of L1688 has been estimated to range between $550\,\sm$ 
(from C$^{18}$O measurements -- Wilking \& Lada \cite{wl83}) and 
$1500\,\sm$ (from CS(2-1) data -- Liseau et al. \cite{liseau95}).
Using $\ms^\mathrm{L1688}= 92\,\sm$, we thus get 
SFE$^\mathrm{L1688} \sim 6-14$~\%, which is somewhat lower than 
previous estimates ($\geq 22\% $ -- WLY89). 
Note, however, that active star formation in L1688 appears to be limited to 
the three sub-clusters/dense cores Oph~A, Oph~EF, and Oph~B (see Fig.~1 and 
Loren et al. \cite{loren90}), where the local star formation efficiency 
is significantly higher: $SFE \sim 31\% $, using a total core mass of 
$200\,\sm$ (Loren et al. \cite{loren90}).

\begin{figure}[ht]
\centering
\includegraphics[width=8.0cm]{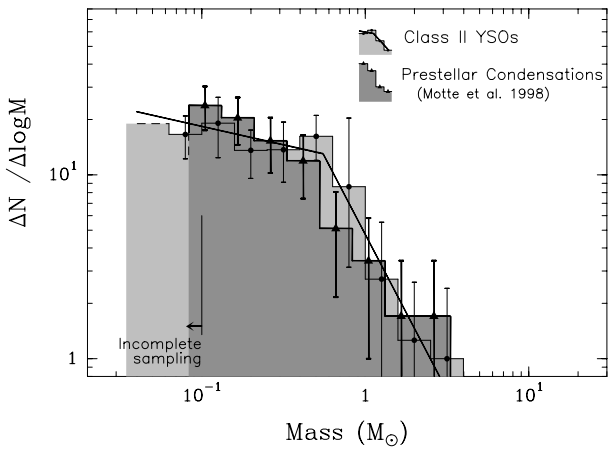}
\caption[]{Comparison of the pre-stellar mass spectrum 
measured by MAN98 for 58 protocluster condensations 
(dark histogram with statistical error bars) with the YSO mass 
function derived in Sect.~5.1 for 123 Class~II systems (light histogram
with error bars accounting for both statistical and age uncertainties). 
}
\label{clump}
\end{figure}
 
\subsection{Comparison with the $\roph$ 
protocluster condensations}

In an extensive 1.3$\,$mm dust continuum imaging survey of L1688 
with the IRAM 30~m 
telescope (11$^{\prime\prime}$ resolution), 
MAN98 could identify 58 
compact starless condensations.
Molecular line observations (e.g. Belloche, 
Andr\'e, \& Motte \cite{bam}) indicate that the condensations are 
gravitationally bound and thus likely pre-stellar in nature. 
MAN98 noted a remarkable similarity between the mass spectrum of these 
pre-stellar  condensations and the IMF of Miller \& Scalo (\cite{ms79}).

In Fig.~10, we compare the pre-stellar mass spectrum determined 
by MAN98 with the mass distribution of Class~II YSOs
derived in Sect. 5.1. (As such, both distributions are uncorrected for the 
presence of close binary systems.)
It can be seen that there is a good agreement 
in shape between the two mass spectra.
A Kolmogorov-Smirnov test performed on the corresponding 
cumulative distributions confirms that they are statistically 
indistinguishable at the 95\% confidence level.  
This supports the suggestion of MAN98 that the IMF of 
embedded clusters is primarily determined by cloud fragmentation at the 
pre-stellar stage of star formation. 
The fact that both the pre-stellar and the YSO spectrum 
of Fig.~10 present a break at roughly the same mass $\sim 0.5\, \sm$ 
is quite remarkable. A small, global shift of the masses by only 
$\sim $~30\% upward or downward in one of the spectra would make 
them differ at the 2$\sigma $ statistical level. Although in absolute 
terms, 
both sets of masses are probably uncertain by a factor of $\sim 2 $ 
(due to uncertainties in the 1.3$\,$mm dust opacity and in the cluster
age, respectively), this suggests that the protocluster 
condensations identified at 1.3$\,$mm may form stars/systems with an 
efficiency larger than $\sim $~50--70\%.

\section{Conclusions}

We have used ISOCAM to survey the $\roph$ main cloud for embedded YSOs 
down to a completeness level of $\sim 10 - 15\,$mJy at 6.7$\,\mu$m 
and 14.3$\,\mu$m. Our main findings are as follows: 
\begin{enumerate}

\item A total of 425 point sources are detected in the 
$\sim 0.7$~deg$^2$ field covered by the survey, of which 211 are seen at 
both 6.7$\,\mu$m and 14.3$\,\mu$m.
The observed distribution of the flux density ratio $\flwt/\flwd$ is clearly 
bimodal with a gap separating two distinct groups of sources: a ``red'' group
corresponding to Class~I and Class~II YSOs with optically thick mid-IR 
excesses, and a ``blue'' group consisting of Class~III YSOs and 
background stars (cf. Fig.~3).

\item The red group
consists of 139 cluster members, 
71 of them being newly identified YSOs. 
Based on their mid-IR colors from 2$\,\mu$m
to $6.7\,\mu$m and $14.3\,\mu$m, essentially all of these new YSOs 
are low-luminosity Class~II objects. This brings the total number of 
Class~II members to 123, a factor of 2 larger than previously known.
Only 50~\% of these Class~II sources have a near-IR excess large enough to 
be recognizable in a near-IR color-color diagram.

\item Combining near-IR data from Barsony et al. (\cite{bklt})
with our mid-IR photometry, we derive stellar luminosities 
for 123 Class~II and 74 Class~III YSOs. We also estimate bolometric 
luminosities for 16 Class~I objects. The corresponding luminosity functions
(Fig.~6) are complete down to $\sim 0.02\,\sol$, $\sim 0.03\,\sol$, 
and $\sim 0.2\,\sol$ for Class~I, Class~II, and Class~III objects, 
respectively.
The luminosity function of Class~II YSOs is essentially flat 
(in logarithmic units) below $\sim 2\,\sol$, with a possible 
peak at $1.5\,\sol$ and dip at $0.5\,\sol$ seen in the 
four sub-regions Oph~A, Oph~B, Oph~EF, and L1689S (Fig.~9).

\item The large proportion of Class~II objects observed above the completeness 
level for Class~III sources ($\ms \sim 0.2\,\sm $) suggests that 
more than 50\% of the $\roph$ embedded YSO population 
have optically thick circumstellar disks at mid-IR wavelengths.
In a majority of cases, the luminosities of these disks are consistent 
with pure reprocessing of stellar light. Only $\sim 35$\% of the Class~II
objects have excess mid-IR luminosities suggestive of substantial disk
accretion rates.

\item 
The luminosity function of Class~II objects is well modeled 
by a population of PMS stars with ages in the range $\sim $~0.3$-$2~Myr  
and a roughly flat mass distribution 
below $\mflat \sim 0.5\,\sm$, i.e., 
{\small d$N/$d$\,$log$\ms \propto \ms^{-0.15}$} down to $\sim 0.06\,\sm$.

\item Pending the results of a deep X-ray census of Class~III objects, 
we argue that the mass distribution of Class~II YSOs is representative 
of the emergent IMF of the embedded cluster.
If we account for the presence of unresolved binaries, this emergent 
mass function is well described by a two-segment power law with  
a low-mass index $\alpha_1 = -0.35 \pm 0.25$, a high-mass index 
$\alpha_2 = -1.7$, and a transition mass $\mflat=0.55 \pm 0.25\,\sm$
(where {\small d$N/$d$\,$log$\ms \propto \ms^{\alpha}$}). 
We find no evidence for a sharp turnover at low masses 
down to at least $\sim 0.06\, \sm$.

\item The shape of the mass function for Class~II systems is 
statistically indistinguishable from the mass 
spectrum determined at 1.3$\,$mm by Motte, Andr\'e, \& Neri (\cite{man}) for the 
pre-stellar condensations of the protocluster. 
This supports the 
conclusion of these authors that the IMF may be primarily determined 
by fragmentation at the pre-stellar stage of star formation. It also suggests
that the 1.3$\,$mm protocluster condensations should form stars/systems 
with an efficiency larger than $\sim $~50--70\%.

\end{enumerate}

\begin{acknowledgements}
      S.B. was supported by an ESA Research Fellowship during his stay at the
      Stockholm Observatory. The authors thank the referee, Andrea Moneti,
      for constructive criticisms.
\end{acknowledgements}

   \begin{table*}
      \caption[]{List of the 212 ISOCAM sources identified as 
      members of the $\roph$ cluster}
         \label{TabId}
      \[
         \begin{tabular}{ccccccclc}
            \hline
            \noalign{\smallskip}
ISO & \multicolumn {2} {c} {Coordinates(J2000)$^a$}
& $\flwd$ & $\sigma_{6.7}\,\,^b$ & $\flwt$ & $\sigma_{14.3}\,\,^b$
& Identification$^c$  & IR$^d$ \\
 
 \#    &    $\alpha$    &    $\delta$  
&   [Jy]  &  [Jy]          & [Jy]    &  [Jy]            
&                 & Class \\
            \noalign{\smallskip}
            \hline
            \noalign{\smallskip}
   1 & 16$^\mathrm{h}$25$^\mathrm{m}$36$\fs$8 & $-$24$\degr$15$\arcmin$51$\arcsec$ &  0.157 & 0.031 &  0.249 & 0.023 & IRS2           &  II  \\
   2 & 16$^\mathrm{h}$25$^\mathrm{m}$38$\fs$2 & $-$24$\degr$22$\arcmin$40$\arcsec$ &  0.075 & 0.014 &  0.088 & 0.007 & B162538$-$242238 & nII  \\
   3 & 16$^\mathrm{h}$25$^\mathrm{m}$39$\fs$7 & $-$24$\degr$26$\arcmin$37$\arcsec$ &  0.197 & 0.020 &  0.449 & 0.013 & IRS3           & nII  \\
   4 & 16$^\mathrm{h}$25$^\mathrm{m}$41$\fs$1 & $-$24$\degr$21$\arcmin$38$\arcsec$ &  0.008 & 0.008 &    -   &   -   & B162541$-$242138 & III  \\
   5 & 16$^\mathrm{h}$25$^\mathrm{m}$50$\fs$6 & $-$24$\degr$39$\arcmin$18$\arcsec$ &  0.045 & 0.012 &    -   &   -   & IRS10          & III  \\
   6 & 16$^\mathrm{h}$25$^\mathrm{m}$56$\fs$3 & $-$24$\degr$20$\arcmin$50$\arcsec$ &  0.400 & 0.042 &  0.662 & 0.024 & SR4/IRS12      &  II  \\
   7 & 16$^\mathrm{h}$25$^\mathrm{m}$57$\fs$5 & $-$24$\degr$30$\arcmin$34$\arcsec$ &  0.049 & 0.007 &  0.009 & 0.006 & GSS20          & III  \\
   8 & 16$^\mathrm{h}$26$^\mathrm{m}$01$\fs$5 & $-$24$\degr$29$\arcmin$47$\arcsec$ &  0.034 & 0.010 &    -   &   -   & B162601$-$242945 & III  \\
   9 & 16$^\mathrm{h}$26$^\mathrm{m}$01$\fs$5 & $-$24$\degr$25$\arcmin$20$\arcsec$ &  0.025 & 0.009 &  0.032 & 0.007 & SKS1$-$4         & nII  \\
  10 & 16$^\mathrm{h}$26$^\mathrm{m}$03$\fs$2 & $-$24$\degr$23$\arcmin$40$\arcsec$ &  0.507 & 0.017 &  0.341 & 0.024 & DoAr21/GSS23   & III  \\
  11 & 16$^\mathrm{h}$26$^\mathrm{m}$03$\fs$2 & $-$24$\degr$17$\arcmin$49$\arcsec$ &  0.025 & 0.002 &  0.006 & 0.003 & VSSG19         & III  \\
  12 & 16$^\mathrm{h}$26$^\mathrm{m}$04$\fs$6 & $-$24$\degr$17$\arcmin$52$\arcsec$ &  0.005 & 0.002 &  0.012 & 0.003 & B162604$-$241753 & nII  \\
  13 & 16$^\mathrm{h}$26$^\mathrm{m}$07$\fs$0 & $-$24$\degr$27$\arcmin$28$\arcsec$ &  0.112 & 0.022 &  0.128 & 0.015 & B162607$-$242725 & nII  \\
  14 & 16$^\mathrm{h}$26$^\mathrm{m}$07$\fs$5 & $-$24$\degr$27$\arcmin$43$\arcsec$ &  0.032 & 0.007 &    -   &   -   & B162607$-$242742 & III  \\
  15 & 16$^\mathrm{h}$26$^\mathrm{m}$08$\fs$7 & $-$24$\degr$18$\arcmin$53$\arcsec$ &  0.010 & 0.003 &    -   &   -   & CRBR4          & III? \\
  16 & 16$^\mathrm{h}$26$^\mathrm{m}$09$\fs$7 & $-$24$\degr$34$\arcmin$12$\arcsec$ &  -$^e$ & - &  -$^e$ & - & SR3/GSS25      & III  \\
  17 & 16$^\mathrm{h}$26$^\mathrm{m}$10$\fs$4 & $-$24$\degr$20$\arcmin$58$\arcsec$ &  0.386 & 0.011 &  0.43  & 0.017 & GSS26          &  II  \\
  18 & 16$^\mathrm{h}$26$^\mathrm{m}$15$\fs$8 & $-$24$\degr$19$\arcmin$23$\arcsec$ &  0.026 & 0.002 &  0.005 & 0.003 & SKS1$-$7         & III? \\
  19 & 16$^\mathrm{h}$26$^\mathrm{m}$17$\fs$0 & $-$24$\degr$22$\arcmin$26$\arcsec$ &  0.262 & 0.005 &  0.414 & 0.012 & GSS29          &  II  \\
  20 & 16$^\mathrm{h}$26$^\mathrm{m}$17$\fs$2 & $-$24$\degr$20$\arcmin$23$\arcsec$ &  0.146 & 0.003 &  0.293 & 0.004 & DoAr24/GSS28   &  II  \\
  21 & 16$^\mathrm{h}$26$^\mathrm{m}$17$\fs$3 & $-$24$\degr$23$\arcmin$49$\arcsec$ &  0.246 & 0.004 &  0.363 & 0.004 & CRBR12         & nI   \\
  22 & 16$^\mathrm{h}$26$^\mathrm{m}$18$\fs$6 & $-$24$\degr$17$\arcmin$11$\arcsec$ &  0.010 & 0.001 &    -   &   -   & B162618$-$241712 & III? \\
  23 & 16$^\mathrm{h}$26$^\mathrm{m}$18$\fs$8 & $-$24$\degr$26$\arcmin$13$\arcsec$ &  0.033 & 0.008 &  0.027 & 0.007 & SKS1$-$10        & nII  \\
  24 & 16$^\mathrm{h}$26$^\mathrm{m}$18$\fs$9 & $-$24$\degr$28$\arcmin$22$\arcsec$ &  0.673 & 0.025 &  0.575 & 0.024 & VSSG1          &  II  \\
  25 & 16$^\mathrm{h}$26$^\mathrm{m}$19$\fs$0 & $-$24$\degr$23$\arcmin$07$\arcsec$ &  0.051 & 0.008 &    -   &   -   & CRBR17         & III? \\
  26 & 16$^\mathrm{h}$26$^\mathrm{m}$19$\fs$2 & $-$24$\degr$24$\arcmin$16$\arcsec$ &  0.056 & 0.006 &  0.074 & 0.004 & CRBR15         & nII  \\
  27 & 16$^\mathrm{h}$26$^\mathrm{m}$20$\fs$7 & $-$24$\degr$08$\arcmin$48$\arcsec$ &  0.030 & 0.010 &  0.014 & 0.006 & WSB28          & III  \\
  28 & 16$^\mathrm{h}$26$^\mathrm{m}$21$\fs$0 & $-$24$\degr$15$\arcmin$42$\arcsec$ &  0.037 & 0.004 &  0.013 & 0.005 & B162621$-$241544 & III? \\
  29 & 16$^\mathrm{h}$26$^\mathrm{m}$21$\fs$5 & $-$24$\degr$23$\arcmin$07$\arcsec$ &  4.09  & 0.12  & 19.1   & 0.26  & GSS30/GY6      &  I   \\
  30 & 16$^\mathrm{h}$26$^\mathrm{m}$21$\fs$7 & $-$24$\degr$26$\arcmin$02$\arcsec$ &  0.028 & 0.007 &  0.033 & 0.014 & GY5            & nII  \\
  31 & 16$^\mathrm{h}$26$^\mathrm{m}$21$\fs$8 & $-$24$\degr$22$\arcmin$49$\arcsec$ &  0.066 & 0.013 &  0.119 & 0.025 & LFAM1          &  I   \\
  32 & 16$^\mathrm{h}$26$^\mathrm{m}$22$\fs$2 & $-$24$\degr$44$\arcmin$36$\arcsec$ &  0.027 & 0.009 &  0.028 & 0.012 & GY3            & nII  \\
  33 & 16$^\mathrm{h}$26$^\mathrm{m}$22$\fs$4 & $-$24$\degr$24$\arcmin$08$\arcsec$ &  0.009 & 0.003 &  0.017 & 0.003 & GY11           & nII  \\
  34 & 16$^\mathrm{h}$26$^\mathrm{m}$22$\fs$6 & $-$24$\degr$22$\arcmin$56$\arcsec$ &  0.052 & 0.008 &  0.084 & 0.016 & GY12           & III  \\
  35 & 16$^\mathrm{h}$26$^\mathrm{m}$23$\fs$1 & $-$24$\degr$28$\arcmin$49$\arcsec$ &  0.018 & 0.008 &  0.026 & 0.014 & GY15           & nII  \\
  36 & 16$^\mathrm{h}$26$^\mathrm{m}$23$\fs$4 & $-$24$\degr$21$\arcmin$02$\arcsec$ &  2.59  & 0.068 &  2.56  & 0.066 & GSS31/GY20     &  II  \\
  37 & 16$^\mathrm{h}$26$^\mathrm{m}$23$\fs$7 & $-$24$\degr$24$\arcmin$40$\arcsec$ &  0.395 & 0.005 &  0.405 & 0.007 & LFAM3/GY21     &  II  \\
  38 & 16$^\mathrm{h}$26$^\mathrm{m}$24$\fs$0 & $-$24$\degr$43$\arcmin$09$\arcsec$ &  0.174 & 0.012 &  0.206 & 0.017 & DoAr25/GY17    &  II  \\
  39 & 16$^\mathrm{h}$26$^\mathrm{m}$24$\fs$1 & $-$24$\degr$24$\arcmin$50$\arcsec$ &  2.65  & 0.10  &  2.16  & 0.067 & S2/GY23        &  II  \\
  40 & 16$^\mathrm{h}$26$^\mathrm{m}$24$\fs$2 & $-$24$\degr$16$\arcmin$14$\arcsec$ &  2.73  & 0.027 &  3.02  & 0.030 & EL24           &  II  \\
  41 & 16$^\mathrm{h}$26$^\mathrm{m}$25$\fs$3 & $-$24$\degr$24$\arcmin$47$\arcsec$ &  0.064 & 0.008 &  0.049 & 0.011 & GY29           & nII  \\
  42 & 16$^\mathrm{h}$26$^\mathrm{m}$27$\fs$7 & $-$24$\degr$26$\arcmin$43$\arcsec$ &  0.003 & 0.001 &    -   &   -   & VSSG29/GY37    & III? \\
  43 & 16$^\mathrm{h}$26$^\mathrm{m}$28$\fs$0 & $-$24$\degr$41$\arcmin$51$\arcsec$ &  0.029 & 0.002 &  0.115 & 0.006 & GY33           & nII  \\
  44 & 16$^\mathrm{h}$26$^\mathrm{m}$28$\fs$5 & $-$24$\degr$15$\arcmin$45$\arcsec$ &  0.023 & 0.004 &  0.010 & 0.004 & B162628$-$241543 & III? \\
  45 & 16$^\mathrm{h}$26$^\mathrm{m}$29$\fs$7 & $-$24$\degr$19$\arcmin$08$\arcsec$ &  0.015 & 0.002 &  0.005 & 0.003 & LFAM8/SKS1$-$19  & III  \\
  46 & 16$^\mathrm{h}$26$^\mathrm{m}$30$\fs$5 & $-$24$\degr$22$\arcmin$59$\arcsec$ &  0.347 & 0.094 &  0.303 & 0.065 & VSSG27/GY51    &  II  \\
  47 & 16$^\mathrm{h}$26$^\mathrm{m}$30$\fs$9 & $-$24$\degr$31$\arcmin$07$\arcsec$ &  0.012 & 0.008 &    -   &   -   & IRS14/GY54     & III? \\
  48 & 16$^\mathrm{h}$26$^\mathrm{m}$34$\fs$2 & $-$24$\degr$23$\arcmin$30$\arcsec$ &  -$^e$ & - &  -$^e$ & - & S1/GY70$^f$        & III  \\
  49 & 16$^\mathrm{h}$26$^\mathrm{m}$36$\fs$7 & $-$24$\degr$18$\arcmin$10$\arcsec$ &  0.007 & 0.002 &    -   &   -   & B162636$-$241811 & III? \\
  50 & 16$^\mathrm{h}$26$^\mathrm{m}$36$\fs$8 & $-$24$\degr$19$\arcmin$01$\arcsec$ &  0.007 & 0.003 &    -   &   -   & B162636$-$241902 & III? \\
  51 & 16$^\mathrm{h}$26$^\mathrm{m}$36$\fs$9 & $-$24$\degr$15$\arcmin$54$\arcsec$ &  0.307 & 0.004 &  0.636 & 0.009 & B162636$-$241554 & nII  \\
  52 & 16$^\mathrm{h}$26$^\mathrm{m}$38$\fs$0 & $-$24$\degr$23$\arcmin$01$\arcsec$ &  0.057 & 0.049 &  0.066 & 0.059 & VSSG4/GY81     & nII  \\
  53 & 16$^\mathrm{h}$26$^\mathrm{m}$38$\fs$6 & $-$24$\degr$23$\arcmin$10$\arcsec$ &  0.035 & 0.045 &  0.046 & 0.043 & GY84$^g$           & nII  \\
  54 & 16$^\mathrm{h}$26$^\mathrm{m}$40$\fs$6 & $-$24$\degr$27$\arcmin$16$\arcsec$ &  0.089 & 0.003 &  0.158 & 0.004 & GY91/CRBR42    &  I   \\
  55 & 16$^\mathrm{h}$26$^\mathrm{m}$40$\fs$7 & $-$24$\degr$30$\arcmin$53$\arcsec$ &  0.004 & 0.002 &    -   &   -   & IRS16/GY92     & III? \\

         \end{tabular}
      \]
   \end{table*}
   \begin{table*}
      \[
         \begin{tabular}{ccccccclc}
 
  56 & 16$^\mathrm{h}$26$^\mathrm{m}$41$\fs$6 & $-$24$\degr$40$\arcmin$15$\arcsec$ &  0.060 & 0.005 &  0.045 & 0.006 & WSB37/GY93     &  II  \\
  57 & 16$^\mathrm{h}$26$^\mathrm{m}$41$\fs$8 & $-$24$\degr$18$\arcmin$01$\arcsec$ &  0.011 & 0.002 &    -   &   -   & B162641$-$241801 & III? \\
  58 & 16$^\mathrm{h}$26$^\mathrm{m}$42$\fs$0 & $-$24$\degr$33$\arcmin$24$\arcsec$ &  0.172 & 0.026 &  0.089 & 0.014 & WL8/GY96       & III  \\
  59 & 16$^\mathrm{h}$26$^\mathrm{m}$42$\fs$1 & $-$24$\degr$31$\arcmin$02$\arcsec$ &  0.047 & 0.002 &  0.059 & 0.003 & WL7/GY98       &  II  \\
  60 & 16$^\mathrm{h}$26$^\mathrm{m}$42$\fs$3 & $-$24$\degr$26$\arcmin$26$\arcsec$ &  0.019 & 0.002 &    -   &   -   & GY101          & III  \\
  61 & 16$^\mathrm{h}$26$^\mathrm{m}$42$\fs$3 & $-$24$\degr$26$\arcmin$34$\arcsec$ &  0.012 & 0.001 &    -   &   -   & GY103          & III  \\
  62 & 16$^\mathrm{h}$26$^\mathrm{m}$42$\fs$9 & $-$24$\degr$20$\arcmin$32$\arcsec$ &  0.284 & 0.007 &  0.290 & 0.008 & GSS37/GY110    &  II  \\
  63 & 16$^\mathrm{h}$26$^\mathrm{m}$43$\fs$1 & $-$24$\degr$23$\arcmin$01$\arcsec$ &  0.024 & 0.008 &  0.028 & 0.007 & GY109          & nII  \\
  64 & 16$^\mathrm{h}$26$^\mathrm{m}$43$\fs$8 & $-$24$\degr$16$\arcmin$35$\arcsec$ &  0.033 & 0.002 &  0.009 & 0.003 & VSSG11         & III  \\
  65 & 16$^\mathrm{h}$26$^\mathrm{m}$44$\fs$0 & $-$24$\degr$34$\arcmin$48$\arcsec$ &  1.27  & 0.043 &  2.57  & 0.029 & WL12/GY111     &  I   \\
  66 & 16$^\mathrm{h}$26$^\mathrm{m}$44$\fs$3 & $-$24$\degr$43$\arcmin$18$\arcsec$ &  0.020 & 0.004 &    -   &   -   & GY112          & III  \\
  67 & 16$^\mathrm{h}$26$^\mathrm{m}$45$\fs$1 & $-$24$\degr$23$\arcmin$10$\arcsec$ &  0.291 & 0.013 &  0.307 & 0.012 & GSS39/GY116    &  II  \\
  68 & 16$^\mathrm{h}$26$^\mathrm{m}$46$\fs$5 & $-$24$\degr$12$\arcmin$03$\arcsec$ &  0.352 & 0.006 &  0.240 & 0.004 & VSS27          &  II  \\
  69 & 16$^\mathrm{h}$26$^\mathrm{m}$47$\fs$0 & $-$24$\degr$44$\arcmin$30$\arcsec$ &  0.011 & 0.002 &    -   &   -   & GY122          & III  \\
  70 & 16$^\mathrm{h}$26$^\mathrm{m}$48$\fs$6 & $-$24$\degr$28$\arcmin$39$\arcsec$ &  0.137 & 0.004 &  0.188 & 0.005 & WL2/GY128      &  II  \\
  71 & 16$^\mathrm{h}$26$^\mathrm{m}$48$\fs$7 & $-$24$\degr$26$\arcmin$28$\arcsec$ &  0.006 & 0.002 &    -   &   -   & GY130          & III? \\
  72 & 16$^\mathrm{h}$26$^\mathrm{m}$49$\fs$0 & $-$24$\degr$38$\arcmin$27$\arcsec$ &  0.101 & 0.002 &  0.086 & 0.002 & WL18/GY129     &  II  \\
  73 & 16$^\mathrm{h}$26$^\mathrm{m}$49$\fs$2 & $-$24$\degr$20$\arcmin$05$\arcsec$ &  0.074 & 0.004 &  0.022 & 0.004 & VSSG3/GY135    & III  \\
  74 & 16$^\mathrm{h}$26$^\mathrm{m}$50$\fs$9 & $-$24$\degr$20$\arcmin$52$\arcsec$ &  0.019 & 0.003 &    -   &   -   & IRS20/GY143    & III? \\
  75 & 16$^\mathrm{h}$26$^\mathrm{m}$52$\fs$1 & $-$24$\degr$30$\arcmin$39$\arcsec$ &  0.026 & 0.002 &  0.026 & 0.003 & GY144          & nII  \\
  76 & 16$^\mathrm{h}$26$^\mathrm{m}$53$\fs$6 & $-$24$\degr$32$\arcmin$36$\arcsec$ &  0.022 & 0.003 &  0.017 & 0.004 & GY146          & nII  \\
  77 & 16$^\mathrm{h}$26$^\mathrm{m}$54$\fs$4 & $-$24$\degr$24$\arcmin$38$\arcsec$ &  0.020 & 0.003 &  0.017 & 0.004 & GY152          & nII  \\
  78 & 16$^\mathrm{h}$26$^\mathrm{m}$54$\fs$5 & $-$24$\degr$26$\arcmin$22$\arcsec$ &  0.091 & 0.003 &  0.070 & 0.004 & VSSG5/GY153    &  II  \\
  79 & 16$^\mathrm{h}$26$^\mathrm{m}$54$\fs$9 & $-$24$\degr$27$\arcmin$05$\arcsec$ &  0.013 & 0.003 &  0.013 & 0.004 & GY154          & nII  \\
  80 & 16$^\mathrm{h}$26$^\mathrm{m}$55$\fs$0 & $-$24$\degr$22$\arcmin$31$\arcsec$ &  0.026 & 0.004 &  0.008 & 0.004 & GY156          & III  \\
  81 & 16$^\mathrm{h}$26$^\mathrm{m}$55$\fs$3 & $-$24$\degr$20$\arcmin$29$\arcsec$ &  0.055 & 0.002 &  0.010 & 0.003 & VSSG7/GY157    & III? \\
  82 & 16$^\mathrm{h}$26$^\mathrm{m}$56$\fs$7 & $-$24$\degr$28$\arcmin$38$\arcsec$ &  0.003 & 0.001 &    -   &   -   & GY163          & III? \\
  83 & 16$^\mathrm{h}$26$^\mathrm{m}$57$\fs$3 & $-$24$\degr$14$\arcmin$00$\arcsec$ &  0.114 & 0.005 &  0.084 & 0.004 & B162656$-$241353 & nII  \\
  84 & 16$^\mathrm{h}$26$^\mathrm{m}$57$\fs$4 & $-$24$\degr$35$\arcmin$39$\arcsec$ &  0.009 & 0.003 &  0.010 & 0.005 & WL21/GY164     & nII  \\
  85 & 16$^\mathrm{h}$26$^\mathrm{m}$58$\fs$3 & $-$24$\degr$37$\arcmin$40$\arcsec$ &  0.015 & 0.004 &  0.011 & 0.007 & CRBR51         & nII  \\
  86 & 16$^\mathrm{h}$26$^\mathrm{m}$58$\fs$4 & $-$24$\degr$21$\arcmin$29$\arcsec$ &  0.022 & 0.003 &  0.018 & 0.004 & IRS26/GY171    & nII  \\
  87 & 16$^\mathrm{h}$26$^\mathrm{m}$58$\fs$6 & $-$24$\degr$18$\arcmin$34$\arcsec$ &  0.020 & 0.003 &  0.014 & 0.003 & B162658$-$241836 & nII  \\
  88 & 16$^\mathrm{h}$26$^\mathrm{m}$58$\fs$8 & $-$24$\degr$45$\arcmin$37$\arcsec$ &  1.41  & 0.090 &  2.29  & 0.17  & SR24$^h$           &  II  \\
  89 & 16$^\mathrm{h}$26$^\mathrm{m}$59$\fs$3 & $-$24$\degr$35$\arcmin$57$\arcsec$ &  0.004 & 0.002 &  0.008 & 0.005 & WL14/GY172     & nII  \\
  90 & 16$^\mathrm{h}$26$^\mathrm{m}$59$\fs$3 & $-$24$\degr$35$\arcmin$01$\arcsec$ &  -$^e$ & - &  -$^e$ & - & WL22/GY174     &  II  \\
  91 & 16$^\mathrm{h}$27$^\mathrm{m}$01$\fs$8 & $-$24$\degr$21$\arcmin$37$\arcsec$ &  0.060 & 0.003 &  0.013 & 0.003 & VSSG8/GY181    & III? \\
  92 & 16$^\mathrm{h}$27$^\mathrm{m}$02$\fs$5 & $-$24$\degr$37$\arcmin$30$\arcsec$ &  -$^e$ & - &  -$^e$ & - & WL16/GY182     &  II  \\
  93 & 16$^\mathrm{h}$27$^\mathrm{m}$03$\fs$0 & $-$24$\degr$26$\arcmin$16$\arcsec$ &  0.026 & 0.003 &  0.021 & 0.004 & GY188          & nII  \\
  94 & 16$^\mathrm{h}$27$^\mathrm{m}$03$\fs$7 & $-$24$\degr$20$\arcmin$07$\arcsec$ &  0.008 & 0.002 &  0.007 & 0.003 & B162703$-$242007 & nII  \\
  95 & 16$^\mathrm{h}$27$^\mathrm{m}$04$\fs$1 & $-$24$\degr$28$\arcmin$33$\arcsec$ &  0.103 & 0.006 &  0.164 & 0.004 & WL1/GY192      &  II  \\
  96 & 16$^\mathrm{h}$27$^\mathrm{m}$04$\fs$4 & $-$24$\degr$43$\arcmin$00$\arcsec$ &  0.020 & 0.003 &    -   &   -   & GY193          & III  \\
  97 & 16$^\mathrm{h}$27$^\mathrm{m}$04$\fs$6 & $-$24$\degr$42$\arcmin$15$\arcsec$ &  0.017 & 0.003 &    -   &   -   & GY194          & III  \\
  98 & 16$^\mathrm{h}$27$^\mathrm{m}$04$\fs$6 & $-$24$\degr$27$\arcmin$16$\arcsec$ &  0.046 & 0.006 &  0.066 & 0.008 & GY195          & nII  \\
  99 & 16$^\mathrm{h}$27$^\mathrm{m}$05$\fs$4 & $-$24$\degr$36$\arcmin$31$\arcsec$ &  0.032 & 0.005 &  0.064 & 0.009 & LFAM26/GY197   &  I   \\
 100 & 16$^\mathrm{h}$27$^\mathrm{m}$05$\fs$7 & $-$24$\degr$40$\arcmin$12$\arcsec$ &  0.007 & 0.002 &    -   &   -   & B162705$-$244013 & III? \\
 101 & 16$^\mathrm{h}$27$^\mathrm{m}$06$\fs$0 & $-$24$\degr$26$\arcmin$19$\arcsec$ &  0.025 & 0.002 &    -   &   -   & IRS30/GY203    & III  \\
 102 & 16$^\mathrm{h}$27$^\mathrm{m}$06$\fs$6 & $-$24$\degr$41$\arcmin$48$\arcsec$ &  0.024 & 0.002 &  0.020 & 0.004 & GY204          & nII  \\
 103 & 16$^\mathrm{h}$27$^\mathrm{m}$07$\fs$0 & $-$24$\degr$38$\arcmin$16$\arcsec$ &  0.449 & 0.006 &  0.732 & 0.015 & WL17/GY205     &  II  \\
 104 & 16$^\mathrm{h}$27$^\mathrm{m}$07$\fs$9 & $-$24$\degr$40$\arcmin$27$\arcsec$ &  0.004 & 0.002 &    -   &   -   & GY207          & III? \\
 105 & 16$^\mathrm{h}$27$^\mathrm{m}$09$\fs$2 & $-$24$\degr$34$\arcmin$10$\arcsec$ &  0.174 & 0.005 &  0.172 & 0.008 & WL10/GY211     &  II  \\
 106 & 16$^\mathrm{h}$27$^\mathrm{m}$09$\fs$3 & $-$24$\degr$12$\arcmin$07$\arcsec$ &  0.058 & 0.003 &  0.044 & 0.003 & B162708$-$241204 & nII  \\
 107 & 16$^\mathrm{h}$27$^\mathrm{m}$09$\fs$6 & $-$24$\degr$40$\arcmin$25$\arcsec$ &  0.079 & 0.002 &  0.096 & 0.004 & GY213          & nII  \\
 108 & 16$^\mathrm{h}$27$^\mathrm{m}$09$\fs$6 & $-$24$\degr$37$\arcmin$21$\arcsec$ & 24.0   & 0.66  & 26.2   & 0.50  & EL29/GY214     &  I   \\
 109 & 16$^\mathrm{h}$27$^\mathrm{m}$09$\fs$6 & $-$24$\degr$29$\arcmin$55$\arcsec$ &  0.014 & 0.006 &    -   &   -   & GY215          & III? \\
 110 & 16$^\mathrm{h}$27$^\mathrm{m}$10$\fs$2 & $-$24$\degr$19$\arcmin$16$\arcsec$ &  1.57  & 0.026 &  2.84  & 0.026 & SR21/VSSG23    &  II  \\
 111 & 16$^\mathrm{h}$27$^\mathrm{m}$10$\fs$3 & $-$24$\degr$33$\arcmin$22$\arcsec$ &  0.009 & 0.002 &    -   &   -   & WL9/GY220      & III? \\
 112 & 16$^\mathrm{h}$27$^\mathrm{m}$11$\fs$4 & $-$24$\degr$40$\arcmin$46$\arcsec$ &  0.221 & 0.006 &  0.391 & 0.010 & GY224          &  II  \\
 113 & 16$^\mathrm{h}$27$^\mathrm{m}$11$\fs$7 & $-$24$\degr$23$\arcmin$49$\arcsec$ &  0.034 & 0.010 &    -   &   -   & IRS32/GY228    & III? \\
 114 & 16$^\mathrm{h}$27$^\mathrm{m}$11$\fs$9 & $-$24$\degr$38$\arcmin$31$\arcsec$ &  0.230 & 0.008 &  0.126 & 0.012 & WL19/GY227     & II/III?  \\

         \end{tabular}
      \]
   \end{table*}
   \begin{table*}
      \[
         \begin{tabular}{ccccccclc}
 115 & 16$^\mathrm{h}$27$^\mathrm{m}$12$\fs$1 & $-$24$\degr$34$\arcmin$48$\arcsec$ &  0.024 & 0.003 &  0.019 & 0.006 & WL11/GY229     & nII  \\
 116 & 16$^\mathrm{h}$27$^\mathrm{m}$13$\fs$9 & $-$24$\degr$18$\arcmin$25$\arcsec$ &  0.074 & 0.005 &  0.073 & 0.005 & B162713$-$241818 & nII  \\
 117 & 16$^\mathrm{h}$27$^\mathrm{m}$14$\fs$0 & $-$24$\degr$43$\arcmin$31$\arcsec$ &  0.093 & 0.003 &  0.168 & 0.006 & GY235          & nII  \\
 118 & 16$^\mathrm{h}$27$^\mathrm{m}$14$\fs$6 & $-$24$\degr$26$\arcmin$55$\arcsec$ &  0.119 & 0.009 &  0.101 & 0.009 & IRS33/GY236    & nII  \\
 119 & 16$^\mathrm{h}$27$^\mathrm{m}$15$\fs$4 & $-$24$\degr$30$\arcmin$54$\arcsec$ &  0.054 & 0.005 &  0.061 & 0.007 & IRS35/GY238    & nII  \\
 120 & 16$^\mathrm{h}$27$^\mathrm{m}$15$\fs$7 & $-$24$\degr$26$\arcmin$46$\arcsec$ &  0.167 & 0.010 &  0.152 & 0.014 & IRS34/GY239    &  II  \\
 121 & 16$^\mathrm{h}$27$^\mathrm{m}$15$\fs$9 & $-$24$\degr$38$\arcmin$46$\arcsec$ &  0.151 & 0.004 &  0.782 & 0.010 & WL20/GY240     &  II  \\
 122 & 16$^\mathrm{h}$27$^\mathrm{m}$16$\fs$1 & $-$24$\degr$25$\arcmin$22$\arcsec$ &  0.018 & 0.010 &  0.019 & 0.005 & IRS36/GY241    & nII  \\
 123$^\star$ & 16$^\mathrm{h}$27$^\mathrm{m}$17$\fs$6 & $-$24$\degr$05$\arcmin$19$\arcsec$ &  0.021 & 0.002 &  0.035 & 0.004 & ISO1627176$-$240519 & nII  \\
 124 & 16$^\mathrm{h}$27$^\mathrm{m}$17$\fs$6 & $-$24$\degr$28$\arcmin$58$\arcsec$ &  0.286 & 0.011 &  0.347 & 0.009 & IRS37/GY244    &  II  \\
 125 & 16$^\mathrm{h}$27$^\mathrm{m}$18$\fs$0 & $-$24$\degr$28$\arcmin$55$\arcsec$ &  0.186 & 0.014 &  0.049 & 0.017 & WL5/GY246      & III  \\
 126 & 16$^\mathrm{h}$27$^\mathrm{m}$18$\fs$0 & $-$24$\degr$24$\arcmin$31$\arcsec$ &  0.028 & 0.004 &    -   &   -   & GY248          & III  \\
 127 & 16$^\mathrm{h}$27$^\mathrm{m}$18$\fs$5 & $-$24$\degr$39$\arcmin$15$\arcsec$ &  0.075 & 0.004 &  0.096 & 0.009 & GY245          &  II  \\
 128 & 16$^\mathrm{h}$27$^\mathrm{m}$18$\fs$6 & $-$24$\degr$29$\arcmin$07$\arcsec$ &  0.284 & 0.019 &  0.272 & 0.012 & WL4/GY247      &  II  \\
 129 & 16$^\mathrm{h}$27$^\mathrm{m}$19$\fs$3 & $-$24$\degr$28$\arcmin$45$\arcsec$ &  0.285 & 0.015 &  0.218 & 0.006 & WL3/GY249      &  II  \\
 130 & 16$^\mathrm{h}$27$^\mathrm{m}$19$\fs$5 & $-$24$\degr$41$\arcmin$43$\arcsec$ &  0.053 & 0.004 &    -   &   -   & SR12/GY250     & III  \\
 131 & 16$^\mathrm{h}$27$^\mathrm{m}$21$\fs$6 & $-$24$\degr$21$\arcmin$52$\arcsec$ &  0.014 & 0.004 &    -   &   -   & GY255          & III? \\
 132 & 16$^\mathrm{h}$27$^\mathrm{m}$21$\fs$7 & $-$24$\degr$41$\arcmin$42$\arcsec$ &  2.68  & 0.096 &  2.13  & 0.057 & IRS42/GY252    &  II  \\
 133 & 16$^\mathrm{h}$27$^\mathrm{m}$21$\fs$8 & $-$24$\degr$43$\arcmin$34$\arcsec$ &  0.035 & 0.002 &  0.008 & 0.004 & GY253          & III  \\
 134 & 16$^\mathrm{h}$27$^\mathrm{m}$21$\fs$8 & $-$24$\degr$29$\arcmin$55$\arcsec$ &  1.09  & 0.029 &  1.26  & 0.033 & WL6/GY254      &  I   \\
 135 & 16$^\mathrm{h}$27$^\mathrm{m}$22$\fs$9 & $-$24$\degr$18$\arcmin$00$\arcsec$ &  0.049 & 0.003 &  0.011 & 0.003 & VSSG22         & III  \\
 136 & 16$^\mathrm{h}$27$^\mathrm{m}$24$\fs$3 & $-$24$\degr$41$\arcmin$46$\arcsec$ &  0.008 & 0.002 &    -   &   -   & GY258          & III? \\
 137 & 16$^\mathrm{h}$27$^\mathrm{m}$24$\fs$8 & $-$24$\degr$41$\arcmin$03$\arcsec$ &  0.177 & 0.006 &  0.362 & 0.015 & CRBR85         &  I   \\
 138 & 16$^\mathrm{h}$27$^\mathrm{m}$26$\fs$0 & $-$24$\degr$19$\arcmin$28$\arcsec$ &  0.006 & 0.002 &  0.006 & 0.002 & B162726$-$241925 & nII  \\
 139 & 16$^\mathrm{h}$27$^\mathrm{m}$26$\fs$2 & $-$24$\degr$42$\arcmin$45$\arcsec$ &  0.032 & 0.003 &  0.039 & 0.006 & GY260          & nII  \\
 140 & 16$^\mathrm{h}$27$^\mathrm{m}$26$\fs$8 & $-$24$\degr$39$\arcmin$22$\arcsec$ &  0.160 & 0.005 &  0.197 & 0.019 & GY262          &  II  \\
 141 & 16$^\mathrm{h}$27$^\mathrm{m}$27$\fs$1 & $-$24$\degr$40$\arcmin$51$\arcsec$ &  2.21  & 0.019 &  4.43  & 0.091 & IRS43/GY265    &  I   \\
 142 & 16$^\mathrm{h}$27$^\mathrm{m}$27$\fs$5 & $-$24$\degr$31$\arcmin$18$\arcsec$ &  0.095 & 0.004 &  0.098 & 0.011 & VSSG25/GY267   &  II  \\
 143 & 16$^\mathrm{h}$27$^\mathrm{m}$28$\fs$3 & $-$24$\degr$39$\arcmin$33$\arcsec$ &  4.19  & 0.14  & 11.4   & 0.12  & IRS44/GY269    &  I   \\
 144 & 16$^\mathrm{h}$27$^\mathrm{m}$28$\fs$4 & $-$24$\degr$27$\arcmin$21$\arcsec$ &  0.407 & 0.016 &  0.481 & 0.018 & IRS45/GY273    &  II  \\
 145 & 16$^\mathrm{h}$27$^\mathrm{m}$29$\fs$7 & $-$24$\degr$39$\arcmin$16$\arcsec$ &  0.294 & 0.004 &  0.649 & 0.011 & IRS46/GY274    &  I   \\
 146 & 16$^\mathrm{h}$27$^\mathrm{m}$29$\fs$7 & $-$24$\degr$33$\arcmin$36$\arcsec$ &  0.013 & 0.002 &    -   &   -   & GY278          & III? \\
 147 & 16$^\mathrm{h}$27$^\mathrm{m}$30$\fs$1 & $-$24$\degr$27$\arcmin$43$\arcsec$ &  1.78  & 0.017 &  1.54  & 0.017 & IRS47/GY279    &  II  \\
 148 & 16$^\mathrm{h}$27$^\mathrm{m}$30$\fs$8 & $-$24$\degr$34$\arcmin$04$\arcsec$ &  0.014 & 0.002 &    -   &   -   & GY283          & III? \\
 149 & 16$^\mathrm{h}$27$^\mathrm{m}$30$\fs$9 & $-$24$\degr$47$\arcmin$25$\arcsec$ &  0.029 & 0.002 &  0.012 & 0.006 & B162730$-$244726 & III  \\
 150$^\star$ & 16$^\mathrm{h}$27$^\mathrm{m}$30$\fs$9 & $-$24$\degr$27$\arcmin$34$\arcsec$ &  0.078 & 0.008 &  0.094 & 0.007 & ISO1627309$-$242734 & nII  \\
 151 & 16$^\mathrm{h}$27$^\mathrm{m}$31$\fs$0 & $-$24$\degr$25$\arcmin$00$\arcsec$ &  0.059 & 0.007 &  0.098 & 0.008 & GY284          & nII  \\
 152 & 16$^\mathrm{h}$27$^\mathrm{m}$32$\fs$5 & $-$24$\degr$33$\arcmin$22$\arcsec$ &  0.030 & 0.002 &  0.018 & 0.004 & GY289          & III  \\
 153 & 16$^\mathrm{h}$27$^\mathrm{m}$32$\fs$6 & $-$24$\degr$32$\arcmin$42$\arcsec$ &  0.020 & 0.003 &    -   &   -   & GY290          & III? \\
 154 & 16$^\mathrm{h}$27$^\mathrm{m}$32$\fs$8 & $-$24$\degr$32$\arcmin$34$\arcsec$ &  0.058 & 0.002 &  0.061 & 0.006 & GY291          & nII  \\
 155 & 16$^\mathrm{h}$27$^\mathrm{m}$33$\fs$2 & $-$24$\degr$41$\arcmin$15$\arcsec$ &  0.391 & 0.006 &  0.534 & 0.013 & GY292          &  II  \\
 156 & 16$^\mathrm{h}$27$^\mathrm{m}$35$\fs$2 & $-$24$\degr$38$\arcmin$31$\arcsec$ &  0.017 & 0.002 &    -   &   -   & GY295          & III? \\
 157 & 16$^\mathrm{h}$27$^\mathrm{m}$35$\fs$6 & $-$24$\degr$45$\arcmin$31$\arcsec$ &  0.010 & 0.002 &    -   &   -   & GY296          & III  \\
 158 & 16$^\mathrm{h}$27$^\mathrm{m}$36$\fs$3 & $-$24$\degr$28$\arcmin$34$\arcsec$ &  0.006 & 0.003 &    -   &   -   & GY297          & III? \\
 159 & 16$^\mathrm{h}$27$^\mathrm{m}$37$\fs$2 & $-$24$\degr$30$\arcmin$34$\arcsec$ &  5.87  & 0.11  &  6.49  & 0.099 & IRS48/GY304    &  I   \\
 160 & 16$^\mathrm{h}$27$^\mathrm{m}$37$\fs$4 & $-$24$\degr$17$\arcmin$58$\arcsec$ &  0.011 & 0.002 &  0.010 & 0.003 & B162737$-$241756 & nII  \\
 161 & 16$^\mathrm{h}$27$^\mathrm{m}$37$\fs$6 & $-$24$\degr$42$\arcmin$40$\arcsec$ &  0.071 & 0.003 &  0.161 & 0.008 & GY301          & nII  \\
 162 & 16$^\mathrm{h}$27$^\mathrm{m}$38$\fs$0 & $-$24$\degr$25$\arcmin$25$\arcsec$ &  0.005 & 0.003 &    -   &   -   & GY309          & III? \\
 163 & 16$^\mathrm{h}$27$^\mathrm{m}$38$\fs$4 & $-$24$\degr$36$\arcmin$58$\arcsec$ &  0.303 & 0.006 &  0.332 & 0.010 & IRS49/GY308    &  II  \\
 164 & 16$^\mathrm{h}$27$^\mathrm{m}$38$\fs$8 & $-$24$\degr$38$\arcmin$39$\arcsec$ &  0.020 & 0.002 &  0.039 & 0.006 & GY310          & nII  \\
 165 & 16$^\mathrm{h}$27$^\mathrm{m}$39$\fs$1 & $-$24$\degr$40$\arcmin$21$\arcsec$ &  0.049 & 0.002 &  0.077 & 0.008 & GY312          & nII  \\
 166 & 16$^\mathrm{h}$27$^\mathrm{m}$39$\fs$5 & $-$24$\degr$39$\arcmin$16$\arcsec$ &  0.266 & 0.005 &  0.411 & 0.008 & GY314          &  II  \\
 167 & 16$^\mathrm{h}$27$^\mathrm{m}$40$\fs$0 & $-$24$\degr$43$\arcmin$13$\arcsec$ &  0.924 & 0.016 &  1.06  & 0.014 & IRS51/GY315    &  I   \\
 168 & 16$^\mathrm{h}$27$^\mathrm{m}$40$\fs$5 & $-$24$\degr$22$\arcmin$07$\arcsec$ &  0.464 & 0.009 &  0.449 & 0.007 & SR9/IRS52      &  II  \\
 169 & 16$^\mathrm{h}$27$^\mathrm{m}$41$\fs$5 & $-$24$\degr$35$\arcmin$37$\arcsec$ &  0.024 & 0.002 &    -   &   -   & GY322          & III? \\
 170 & 16$^\mathrm{h}$27$^\mathrm{m}$41$\fs$8 & $-$24$\degr$46$\arcmin$45$\arcsec$ &  0.026 & 0.002 &  0.043 & 0.006 & B162741$-$244645 & nII  \\
 171 & 16$^\mathrm{h}$27$^\mathrm{m}$41$\fs$9 & $-$24$\degr$43$\arcmin$37$\arcsec$ &  0.050 & 0.002 &  0.058 & 0.005 & GY323          & nII  \\
 172 & 16$^\mathrm{h}$27$^\mathrm{m}$42$\fs$7 & $-$24$\degr$38$\arcmin$49$\arcsec$ &  0.021 & 0.004 &  0.013 & 0.007 & GY326          & nII  \\
 173 & 16$^\mathrm{h}$27$^\mathrm{m}$43$\fs$7 & $-$24$\degr$43$\arcmin$07$\arcsec$ &  0.027 & 0.003 &    -   &   -   & IRS53/GY334    & III? \\

         \end{tabular}
      \]
   \end{table*}
   \begin{table*}
      \[
         \begin{tabular}{ccccccclc}

 174 & 16$^\mathrm{h}$27$^\mathrm{m}$45$\fs$9 & $-$24$\degr$38$\arcmin$00$\arcsec$ &  0.003 & 0.002 &    -   &   -   & GY346          & III? \\
 175 & 16$^\mathrm{h}$27$^\mathrm{m}$46$\fs$0 & $-$24$\degr$44$\arcmin$52$\arcsec$ &  0.061 & 0.002 &  0.082 & 0.004 & GY344          & nII  \\
 176 & 16$^\mathrm{h}$27$^\mathrm{m}$46$\fs$2 & $-$24$\degr$31$\arcmin$40$\arcsec$ &  0.024 & 0.003 &  0.022 & 0.005 & GY350          & nII  \\
 177 & 16$^\mathrm{h}$27$^\mathrm{m}$47$\fs$2 & $-$24$\degr$45$\arcmin$33$\arcsec$ &  0.047 & 0.003 &  0.037 & 0.007 & GY352          & nII  \\
 178 & 16$^\mathrm{h}$27$^\mathrm{m}$49$\fs$9 & $-$24$\degr$25$\arcmin$25$\arcsec$ &  0.055 & 0.005 &  0.041 & 0.005 & GY371          & nII  \\
 179 & 16$^\mathrm{h}$27$^\mathrm{m}$50$\fs$0 & $-$24$\degr$44$\arcmin$15$\arcsec$ &  0.009 & 0.002 &    -   &   -   & GY370          & III? \\
 180 & 16$^\mathrm{h}$27$^\mathrm{m}$50$\fs$2 & $-$24$\degr$25$\arcmin$43$\arcsec$ &  0.182 & 0.005 &  0.045 & 0.006 & VSSG14/GY372   & III  \\
 181 & 16$^\mathrm{h}$27$^\mathrm{m}$50$\fs$3 & $-$24$\degr$39$\arcmin$01$\arcsec$ &  0.004 & 0.002 &    -   &   -   & GY373          & III? \\
 182 & 16$^\mathrm{h}$27$^\mathrm{m}$51$\fs$7 & $-$24$\degr$31$\arcmin$46$\arcsec$ &  4.65  & 0.094 &  5.3   & 0.11  & IRS54/GY378    &  I   \\
 183 & 16$^\mathrm{h}$27$^\mathrm{m}$51$\fs$9 & $-$24$\degr$46$\arcmin$28$\arcsec$ &  0.015 & 0.002 &    -   &   -   & GY377          & III  \\
 184 & 16$^\mathrm{h}$27$^\mathrm{m}$52$\fs$0 & $-$24$\degr$40$\arcmin$48$\arcsec$ &  0.067 & 0.002 &  0.017 & 0.004 & IRS55/GY380    & III  \\
 185 & 16$^\mathrm{h}$27$^\mathrm{m}$55$\fs$0 & $-$24$\degr$28$\arcmin$39$\arcsec$ &  0.025 & 0.002 &  0.026 & 0.002 & GY397          & nII  \\
 186 & 16$^\mathrm{h}$27$^\mathrm{m}$55$\fs$5 & $-$24$\degr$44$\arcmin$49$\arcsec$ &  0.011 & 0.002 &    -   &   -   & GY398          & III? \\
 187 & 16$^\mathrm{h}$27$^\mathrm{m}$55$\fs$9 & $-$24$\degr$26$\arcmin$22$\arcsec$ &  0.102 & 0.003 &  0.101 & 0.004 & SR10/GY400     &  II  \\
 188 & 16$^\mathrm{h}$27$^\mathrm{m}$57$\fs$7 & $-$24$\degr$40$\arcmin$01$\arcsec$ &  0.017 & 0.001 &    -   &   -   & GY410          & III  \\
 189 & 16$^\mathrm{h}$27$^\mathrm{m}$57$\fs$8 & $-$24$\degr$36$\arcmin$01$\arcsec$ &  0.004 & 0.003 &    -   &   -   & GY412          & III? \\
 190 & 16$^\mathrm{h}$28$^\mathrm{m}$03$\fs$8 & $-$24$\degr$34$\arcmin$43$\arcsec$ &  0.009 & 0.001 &  0.012 & 0.003 & GY450          & nII  \\
 191 & 16$^\mathrm{h}$28$^\mathrm{m}$04$\fs$9 & $-$24$\degr$34$\arcmin$52$\arcsec$ &  0.036 & 0.002 &  0.012 & 0.003 & GY463          & III? \\
 192 & 16$^\mathrm{h}$28$^\mathrm{m}$05$\fs$5 & $-$24$\degr$33$\arcmin$55$\arcsec$ &  0.002 & 0.001 &    -   &   -   & GY472          & III? \\
 193 & 16$^\mathrm{h}$28$^\mathrm{m}$12$\fs$2 & $-$24$\degr$11$\arcmin$37$\arcsec$ &  0.018 & 0.003 &  0.025 & 0.004 & B162812$-$241138 & nII  \\
 194 & 16$^\mathrm{h}$28$^\mathrm{m}$13$\fs$7 & $-$24$\degr$32$\arcmin$48$\arcsec$ &  0.052 & 0.001 &  0.040 & 0.002 & B162813$-$243249 & nII  \\
 195$^\star$ & 16$^\mathrm{h}$28$^\mathrm{m}$16$\fs$8 & $-$24$\degr$05$\arcmin$19$\arcsec$ &  0.102 & 0.004 &  0.065 & 0.004 & ISO1628168$-$240519 & nII  \\
 196 & 16$^\mathrm{h}$28$^\mathrm{m}$16$\fs$8 & $-$24$\degr$37$\arcmin$04$\arcsec$ &  0.108 & 0.002 &  0.128 & 0.002 & WSB60          &  II  \\
 197 & 16$^\mathrm{h}$28$^\mathrm{m}$22$\fs$1 & $-$24$\degr$42$\arcmin$49$\arcsec$ &  0.017 & 0.001 &  0.010 & 0.002 & B162821$-$244246 & nII  \\
 198 & 16$^\mathrm{h}$28$^\mathrm{m}$33$\fs$2 & $-$24$\degr$22$\arcmin$46$\arcsec$ &  0.701 & 0.020 &  0.265 & 0.004 & SR20           & III  \\
 199 & 16$^\mathrm{h}$28$^\mathrm{m}$45$\fs$6 & $-$24$\degr$28$\arcmin$22$\arcsec$ &  0.315 & 0.004 &  0.230 & 0.006 & SR13           &  II  \\
 200$^\star$ & 16$^\mathrm{h}$31$^\mathrm{m}$43$\fs$4 & $-$24$\degr$55$\arcmin$24$\arcsec$ &  0.170 & 0.003 &  0.430 & 0.006 & ISO1631434$-$245524 & nII  \\
 201$^\star$ & 16$^\mathrm{h}$31$^\mathrm{m}$48$\fs$7 & $-$24$\degr$54$\arcmin$32$\arcsec$ &  0.016 & 0.002 &  0.011 & 0.004 & ISO1631487$-$245432 & nII  \\
 202$^\star$ & 16$^\mathrm{h}$31$^\mathrm{m}$51$\fs$8 & $-$24$\degr$57$\arcmin$26$\arcsec$ &  0.003 & 0.001 &  0.011 & 0.005 & ISO1631518$-$245726 & nII  \\
 203$^\star$ & 16$^\mathrm{h}$31$^\mathrm{m}$52$\fs$0 & $-$24$\degr$55$\arcmin$36$\arcsec$ &  0.080 & 0.002 &  0.322 & 0.004 & ISO1631520$-$245536 & nII  \\
 204 & 16$^\mathrm{h}$31$^\mathrm{m}$52$\fs$0 & $-$24$\degr$56$\arcmin$18$\arcsec$ &  1.19  & 0.011 &  1.84  & 0.016 & L1689$-$IRS5     &  II  \\
 205$^\star$ & 16$^\mathrm{h}$31$^\mathrm{m}$53$\fs$1 & $-$24$\degr$55$\arcmin$04$\arcsec$ &  0.073 & 0.001 &  0.076 & 0.002 & ISO1631531$-$245504 & nII  \\
 206$^\star$ & 16$^\mathrm{h}$31$^\mathrm{m}$54$\fs$8 & $-$25$\degr$03$\arcmin$50$\arcsec$ &  0.055 & 0.002 &  0.045 & 0.004 & ISO1631548$-$250350 & nII  \\
 207 & 16$^\mathrm{h}$31$^\mathrm{m}$55$\fs$2 & $-$25$\degr$03$\arcmin$24$\arcsec$ &  0.355 & 0.005 &  0.582 & 0.006 & IRAS16289$-$2457 & nII  \\
 208$^\star$ & 16$^\mathrm{h}$31$^\mathrm{m}$59$\fs$1 & $-$24$\degr$54$\arcmin$42$\arcsec$ &  0.023 & 0.002 &  0.026 & 0.002 & ISO1631591$-$245442 & nII  \\
 209 & 16$^\mathrm{h}$32$^\mathrm{m}$01$\fs$0 & $-$24$\degr$56$\arcmin$44$\arcsec$ &  0.982 & 0.019 &  1.17  & 0.011 & IRS67/L1689$-$IRS6          &  I   \\
 210$^\star$ & 16$^\mathrm{h}$32$^\mathrm{m}$02$\fs$1 & $-$24$\degr$56$\arcmin$16$\arcsec$ &  0.012 & 0.003 &  0.031 & 0.003 & ISO1632021$-$245616 & nII  \\
 211$^\star$ & 16$^\mathrm{h}$32$^\mathrm{m}$05$\fs$6 & $-$25$\degr$02$\arcmin$34$\arcsec$ &  0.019 & 0.002 &  0.031 & 0.003 & ISO1632056$-$250234 & nII  \\
 212 & 16$^\mathrm{h}$32$^\mathrm{m}$21$\fs$2 & $-$24$\degr$30$\arcmin$35$\arcsec$ &  0.489 & 0.005 &  0.807 & 0.004 & L1689$-$IRS7     &  II  \\
   
            \noalign{\smallskip}
            \hline
         \end{tabular}
      \]
\begin{list}{}{}
\item[ $^{a}$ ]{The absolute ISOCAM 
coordinates have been determined using the Barsony et al. 
(\cite{bklt}) database as a reference frame. The resulting positional 
uncertainty is estimated to be $\sim 3\arcsec$ (rms).}
\item[ $^{b}$ ]{The quoted rms uncertainties on $\flwd$ and $\flwt$ 
comprise the random errors estimated from the temporal and spatial 
noises measured on the data. They do not include the systematic
errors due to absolute calibration and aperture correction uncertainties.
The maximum systematic error is estimated to be less than $\sim 15\,$\%.}
\item[ $^{c}$ ]{For a more complete cross-identification of the $\roph$ IR 
sources, see, e.g., Table~1 of AM94 and Table~3 of
 Barsony et al. (\cite{bklt}). 
``B'' refers to Barsony et al. (\cite{bklt}). ``ISO'' refers to the present
survey. Twelve ISO sources are completely new sources with respect to 
published IR surveys; they are named after their ISOCAM coordinates and 
marked by a $^\star$.}
\item[ $^{d}$ ]{Adopted SED classes for the ISOCAM sources 
(see text in Sect.~3).
``nI'' and ``nII'' indicate newly identified Class~I and Class~II YSOs; 
a question mark indicates an uncertain classification.}
\item[ $^{e}$ ]{ISO16$\,=\,$SR3, ISO48$\,=\,$S1, ISO90$\,=\,$WL22, and ISO92$\,=\,$WL16 
are extended sources in the mid-IR range, most probably due to 
PAH-like emission. No attempt has been made to estimate their integrated  
$\flwd$ and $\flwt$ fluxes.}
\item[ $^{f}$ ]{No precise mid-IR position could be derived for ISO48$\,=\,$S1/GY70.
The quoted coordinates correspond to the near-IR position of 
Barsony et al. (\cite{bklt}).}
\item[ $^{g}$ ]{The position of ISO53 is offset by $\sim 16\arcsec$ to the 
south of the near-IR source GY84. This large mid-IR offset is most likely 
due to the presence of extended, structured emission from the bright 
neighboring source S1.}
\item[ $^{h}$ ]{ISO88$\,=\,$SR24 is a binary system (SR24S/GY167 and 
SR24N/GY168; e.g. Greene et al. \cite{gwayl}) which 
is not resolved in the ISOCAM images.}
\end{list} 
   \end{table*}

   \begin{table*}
      \caption[]{Class~I YSOs}
         \label{tabcl1}
      \[
         \begin{tabular}{clcc}
            \hline
            \noalign{\smallskip}
ISO     & Identification & $\aircl$ &  $\lbol$$^b$  \\
\#$^a$  &                &          &  [$\sol$]     \\
            \noalign{\smallskip}
            \hline
            \noalign{\smallskip}

  182  & IRS54/GY378    &   1.76 &   6.6   \\
  143  & IRS44/GY269    &   1.57 &   8.7   \\
  137  & CRBR85         &   1.48 &   0.36  \\
   99  & LFAM26/GY197   &   1.25 &   0.064 \\
   29  & GSS30/GY6      &   1.20 &  21.    \\
   31  & LFAM1          &   1.08 &   0.13  \\
   65  & WL12/GY111     &   1.04 &   2.6   \\
  108  & EL29/GY214     &   0.98 &  26.    \\
  141  & IRS43/GY265    &   0.98 &   6.7   \\
  145  & IRS46/GY274    &   0.94 &   0.62  \\
   21  & CRBR12         &   0.91 &   0.42  \\
  209  & IRS67/L1689$-$IRS6          &   0.74 &   1.5   \\
   54  & GY91/CRBR42    &   0.70 &   0.17  \\
  134  & WL6/GY254      &   0.59 &   1.7   \\
  159  & IRS48/GY304$^c$    &   0.18 &   7.4   \\
  167  & IRS51/GY315$^c$    &  $-$0.04 &   1.1   \\

            \noalign{\smallskip}
            \hline
         \end{tabular}
      \]
\begin{list}{}{}
\item[ $^{a}$ ]{The ISO number refers to the numbering of Table~1(available 
only in electronic form at http://cdsweb.u-strasbg.fr/).}
\item[ $^{b}$ ]{$\lbol$ is estimated by integrating under the observed SED 
from 1.2$\,\mu$m to 60$\,\mu$m or 100$\,\mu$m and extrapolating from 
60$\,\mu$m or 100$\,\mu$m to 200$\,\mu$m with a spectral index of $\air=-1.0$. 
For the 9 weak sources (ISO$\,$\#~137, 99, 31, 65, 146, 21, 209, 54 and 133) without 
reliable IRAS fluxes, $\lbol$ is 
estimated using a typical $\lbol/\lcal(6.7-14.3\mu$m) 
ratio of 9.8 (see Sect.~4.3).}
\item[ $^{c}$ ]{Although ISO159$\,=\,$IRS48 and ISO167$\,=\,$IRS51 formally lie 
below our practical Class~I$-$Class~II limit ($\aircl = 0.55$), they are 
still considered as Class~I YSOs here (cf. WLY89 and AM94).}
\end{list} 
   \end{table*}

   
   \begin{table*}
      \caption[]{Class~II YSOs}
         \label{tabcl2}
      \[
         \begin{tabular}{clcccccc}
            \hline
            \noalign{\smallskip}
 ISO    & Identification & $\aircl$ & $M_J$ & $M_H$ & $\av$ &  $\ls$   & $\ld$ \\
\#$^a$  &                &          & [mag] & [mag] & [mag] & [$\sol$] & [$\sol$] \\
            \noalign{\smallskip}
            \hline
            \noalign{\smallskip}
	    
  170  & B162741$-$244645 &   0.51 &     -   &   6.1 &  24.9 &   0.018 &   0.047 \\
  103  & WL17/GY205     &   0.42 &   5.2 &     -   &  22.5 &   0.12  &   0.76  \\
  124  & IRS37/GY244    &   0.35 &     -   &   2.4 &  36.9 &   0.99  &   0.50  \\
  112  & GY224          &   0.34 &     -   &   2.4 &  36.2 &   1.1   &   0.56  \\
  118  & IRS33/GY236    &   0.32 &     -   &   3.6 &  38.3 &   0.28  &   0.15  \\
   33  & GY11           &   0.31 &  10.1 &     -   &   2.7 &   0.001 &   0.010 \\
  119  & IRS35/GY238    &   0.30 &     -   &   3.4 &  45.7 &   0.34  &   0.11  \\
  129  & WL3/GY249      &   0.23 &     -   &   2.2 &  42.2 &   1.3   &   0.34  \\
   75  & GY144          &   0.20 &     -   &   5.9 &  26.8 &   0.023 &   0.030 \\
  147  & IRS47/GY279    &   0.17 &   2.6 &     -   &  26.8 &   1.9   &   1.8   \\
   46  & VSSG27/GY51    &   0.17 &   5.2 &     -   &  21.6 &   0.11  &   0.30  \\
  127  & GY245          &   0.17 &   6.7$^b$ &     -   &  24.7 &   0.023 &   0.11  \\
  161  & GY301          &   0.12 &     -   &   2.1 &  44.9 &   1.5   &   0.25  \\
  132  & IRS42/GY252    &   0.08 &   2.1 &     -   &  27.7 &   3.1   &   2.5   \\
   77  & GY152          &   0.05 &     -   &     -   &     -   &   0.037$^c$ &   0.015 \\
   70  & WL2/GY128      &   0.05 &     -   &   2.3 &  38.6 &   1.1   &   0.27  \\
  165  & GY312          &   0.03 &   6.6 &     -   &  14.8 &   0.027 &   0.064 \\
   85  & CRBR51         &   0.03 &     -   &     -   &     -   &   0.025$^c$ &   0.010 \\
  175  & GY344          &   0.02 &   6.5 &     -   &  17.4 &   0.030 &   0.074 \\
   26  & CRBR15         &   0.01 &   6.7 &     -   &  14.6 &   0.022 &   0.061 \\
  139  & GY260          &  $-$0.03 &     -   &   3.7 &  40.3 &   0.24  &   0.058 \\
   37  & LFAM3/GY21     &  $-$0.06 &   4.5 &     -   &  14.5 &   0.25  &   0.33  \\
  121  & WL20/GY240     &  $-$0.07 &   3.3 &     -   &  16.5 &   0.86  &   0.67  \\
   51  & B162636$-$241554 &  $-$0.09 &   4.4 &     -   &   7.6 &   0.28  &   0.43  \\
   95  & WL1/GY192      &  $-$0.11 &   4.9 &     -   &  20.9 &   0.15  &   0.16  \\
  122  & IRS36/GY241    &  $-$0.11 &     -   &   4.9 &  34.3 &   0.066 &   0.025 \\
  
         \end{tabular}
      \]
   \end{table*}
   \begin{table*}
      \[
         \begin{tabular}{clcccccc}
	    
  171  & GY323          &  $-$0.12 &     -   &   4.0 &  30.9 &   0.19  &   0.070 \\
  107  & GY213          &  $-$0.15 &   5.6 &     -   &  19.4 &   0.075 &   0.089 \\
   76  & GY146          &  $-$0.16 &     -   &   4.1 &  43.5 &   0.17  &   0.025 \\
  120  & IRS34/GY239    &  $-$0.23 &     -   &   2.4 &  34.8 &   1.1   &   0.19  \\
  144  & IRS45/GY273    &  $-$0.24 &   3.6 &     -   &  19.5 &   0.66  &   0.45  \\
  204  & L1689$-$IRS5$^d$     &  $-$0.25 &   2.6 &     -   &  12.8 &   1.9   &   1.4   \\
   17  & GSS26          &  $-$0.30 &   3.2 &     -   &  22.6 &   0.95  &   0.43  \\
   93  & GY188          &  $-$0.36 &   6.3 &     -   &  20.5 &   0.034 &   0.020 \\
   98  & GY195          &  $-$0.36 &   5.5 &     -   &  20.6 &   0.087 &   0.063 \\
   53  & GY84           &  $-$0.39 &   5.8 &     -   &  14.1 &   0.060 &   0.037 \\
  190  & GY450          &  $-$0.39 &   9.2 &     -   &   8.6 &   0.002 &   0.008 \\
   23  & SKS1$-$10        &  $-$0.41 &   7.2 &     -   &   8.2 &   0.013 &   0.018 \\
   13  & B162607$-$242725 &  $-$0.41 &   4.6 &     -   &  19.5 &   0.22  &   0.12  \\
  117  & GY235          &  $-$0.43 &   4.9 &     -   &   9.9 &   0.16  &   0.12  \\
  212  & L1689$-$IRS7$^d$     &  $-$0.44 &   2.6 &     -   &  14.7 &   1.8   &   0.64  \\
   79  & GY154          &  $-$0.44 &     -   &   5.5 &  24.3 &   0.033 &   0.013 \\
   24  & VSSG1          &  $-$0.49 &   3.2 &     -   &  17.1 &   0.97  &   0.49  \\
    3  & IRS3           &  $-$0.50 &   4.3 &     -   &   4.6 &   0.31  &   0.28  \\
   39  & S2/GY23        &  $-$0.51 &   2.0 &     -   &  11.8 &   3.7   &   1.6   \\
  140  & GY262          &  $-$0.52 &   3.5 &     -   &  23.7 &   0.71  &   0.20  \\
   40  & EL24           &  $-$0.54 &   1.8 &     -   &  10.0 &   4.5   &   2.1   \\
   52  & VSSG4/GY81     &  $-$0.54 &   5.2 &     -   &  17.6 &   0.11  &   0.058 \\
  154  & GY291          &  $-$0.60 &   4.3 &     -   &  23.0 &   0.29  &   0.058 \\
   94  & B162703$-$242007 &  $-$0.61 &     -   &   6.6 &  16.6 &   0.010 &   0.006 \\
  128  & WL4/GY247      &  $-$0.67 &   3.2 &     -   &  19.5 &   1.1   &   0.24  \\
   59  & WL7/GY98       &  $-$0.69 &   4.1 &     -   &  27.2 &   0.37  &   0.061 \\
   84  & WL21/GY164     &  $-$0.70 &   7.8 &     -   &  14.2 &   0.007 &   0.008 \\
   67  & GSS39/GY116    &  $-$0.72 &   3.1 &     -   &  16.3 &   1.1   &   0.25  \\
   41  & GY29           &  $-$0.73 &   5.1 &     -   &  19.1 &   0.13  &   0.043 \\
   88  & SR24N/GY168    &  $-$0.74 &   2.5 &     -   &   7.7 &   2.1   &   0.76  \\
   35  & GY15           &  $-$0.75 &   6.2 &     -   &  11.5 &   0.040 &   0.019 \\
  164  & GY310          &  $-$0.75 &   6.4 &     -   &   4.0 &   0.032 &   0.024 \\
   63  & GY109          &  $-$0.77 &   5.8 &     -   &  14.6 &   0.064 &   0.022 \\
  151  & GY284          &  $-$0.77 &   4.9 &     -   &   7.6 &   0.15  &   0.065 \\
   43  & GY33           &  $-$0.78 &   4.2 &     -   &  15.5 &   0.36  &   0.092 \\
  138  & B162726$-$241925 &  $-$0.78 &   7.9 &     -   &  11.2 &   0.006 &   0.004 \\
   36  & GSS31/GY20     &  $-$0.79 &   1.5 &     -   &   6.1 &   5.9   &   1.6   \\
  177  & GY352          &  $-$0.79 &   5.2 &     -   &  17.7 &   0.11  &   0.031 \\
  197  & B162821$-$244246 &  $-$0.81 &   6.0 &     -   &  20.3 &   0.050 &   0.009 \\
  110  & SR21/VSSG23    &  $-$0.81 &   1.9 &     -   &   3.5 &   4.0   &   1.7   \\
  166  & GY314          &  $-$0.83 &   3.4 &     -   &   6.4 &   0.80  &   0.26  \\
    9  & SKS1$-$4         &  $-$0.85 &   6.0 &     -   &   9.5 &   0.051 &   0.022 \\
   12  & B162604$-$241753 &  $-$0.86 &   6.6 &     -   &  13.4 &   0.027 &   0.009 \\
  155  & GY292          &  $-$0.90 &   2.7 &     -   &  10.8 &   1.6   &   0.37  \\
   88  & SR24S/GY167    &  $-$0.91 &   2.5 &     -   &   5.9 &   2.2   &   0.72  \\
   19  & GSS29          &  $-$0.91 &   2.9 &     -   &   9.4 &   1.4   &   0.28  \\
  115  & WL11/GY229     &  $-$0.92 &   6.3 &     -   &  13.8 &   0.037 &   0.015 \\
  196  & WSB60$^e$          &  $-$0.92 &   4.9 &     -   &   2.8 &   0.16  &   0.075 \\
   30  & GY5            &  $-$0.92 &   6.3 &     -   &   2.5 &   0.036 &   0.019 \\
  176  & GY350          &  $-$0.94 &   6.4 &     -   &   6.4 &   0.033 &   0.014 \\
   72  & WL18/GY129     &  $-$0.94 &   4.7 &     -   &  10.4 &   0.19  &   0.061 \\
  163  & IRS49/GY309    &  $-$0.96 &   3.1 &     -   &  10.1 &   1.1   &   0.23  \\
  193  & B162812$-$241138 &  $-$0.99 &   6.2 &     -   &   6.2 &   0.039 &   0.016 \\
    2  & B162538$-$242238 &  $-$1.00 &   4.6 &     -   &  10.9 &   0.22  &   0.063 \\
   78  & VSSG5/GY153    &  $-$1.02 &   3.8 &     -   &  19.7 &   0.55  &   0.056 \\
   86  & IRS26/GY171    &  $-$1.04 &   5.2 &     -   &  19.9 &   0.12  &   0.015 \\
  105  & WL10/GY211     &  $-$1.05 &   3.4 &     -   &  12.5 &   0.78  &   0.12  \\
   87  & B162658$-$241836 &  $-$1.06 &   6.0 &     -   &  14.2 &   0.050 &   0.011 \\
  160  & B162737$-$241756 &  $-$1.08 &   7.2 &     -   &   4.5 &   0.014 &   0.006 \\
   83  & B162656$-$241353 &  $-$1.08 &   4.2 &     -   &  11.1 &   0.33  &   0.058 \\

         \end{tabular}
      \]
   \end{table*}
   \begin{table*}
      \[
         \begin{tabular}{clcccccc}
	    
   32  & GY3            &  $-$1.09 &   6.5 &     -   &   0.4 &   0.029 &   0.015 \\
    1  & IRS2           &  $-$1.09 &   3.4 &     -   &   4.9 &   0.79  &   0.15  \\
  185  & GY397          &  $-$1.10 &   6.1 &     -   &   4.1 &   0.045 &   0.016 \\
    6  & SR4/IRS12      &  $-$1.12 &   2.8 &     -   &   1.9 &   1.5   &   0.37  \\
  142  & VSSG25/GY267   &  $-$1.13 &   4.0 &     -   &   9.8 &   0.42  &   0.065 \\
   20  & DoAr24/GSS28   &  $-$1.14 &   3.5 &     -   &   1.8 &   0.72  &   0.16  \\
  178  & GY371          &  $-$1.16 &   5.2 &     -   &   7.1 &   0.11  &   0.026 \\
   62  & GSS37/GY110    &  $-$1.19 &   2.7 &     -   &   8.5 &   1.6   &   0.18  \\
   89  & WL14/GY172     &  $-$1.24 &   5.8 &     -   &  17.2 &   0.059 &   0.006 \\
  194  & B162813$-$243249 &  $-$1.24 &   5.2 &     -   &   5.9 &   0.12  &   0.024 \\
  116  & B162713$-$241818 &  $-$1.27 &   3.7 &     -   &  10.7 &   0.57  &   0.046 \\
  106  & B162708$-$241204 &  $-$1.33 &   4.5 &     -   &   8.0 &   0.24  &   0.027 \\
  102  & GY204          &  $-$1.33 &   6.2 &     -   &   1.5 &   0.039 &   0.011 \\
  168  & SR9/IRS52      &  $-$1.35 &   2.7 &     -   &  $-$0.7 &   1.6   &   0.23  \\
  199  & SR13           &  $-$1.37 &   3.4 &     -   &  $-$0.9 &   0.78  &   0.12  \\
  187  & SR10/GY400     &  $-$1.39 &   4.1 &     -   &  $-$3.2 &   0.38  &   0.052 \\
   68  & VSS27          &  $-$1.53 &   2.6 &     -   &   5.6 &   1.9   &   0.13  \\
  172  & GY326          &  $-$1.56 &   5.3 &     -   &   8.5 &   0.11  &   0.008 \\
   56  & WSB37/GY93     &  $-$1.58 &   4.7 &     -   &   1.2 &   0.20  &   0.024 \\
   38  & DoAr25/GY17    &  $-$1.58 &   3.4 &     -   &   0.7 &   0.83  &   0.11  \\
   -   & GY256          &     -    &     -   &   4.5 &  34.5 &   0.11  &     -   \\
   -   & GY257          &     -    &     -   &   4.1 &  33.2 &   0.17  &     -   \\
   90  & WL22/GY174     &     -    &     -   &     -   &     -   &  29.$^f$    &     -   \\
   92  & WL16/GY182     &     -    &   -     &     -   &     -   &  44.$^f$    &     -   \\
  207  & IRAS16289$-$2457 &     -    &     -   &     -   &     -   &   1.3$^c$   &   0.51  \\
  123  & New$^g$ &     -    &     -   &     -   &     -   &   0.077$^c$ &   0.031 \\
  150  & New$^g$ &     -    &     -   &     -   &     -   &   0.21$^c$  &   0.083 \\
  195  & New$^g$ &     -    &     -   &     -   &     -   &   0.14$^c$  &   0.058 \\
  200  & New$^g$ &     -    &     -   &     -   &     -   &   0.95$^c$  &   0.38  \\
  201  & New$^g$ &     -    &     -   &     -   &     -   &   0.025$^c$ &   0.010 \\
  202  & New$^g$ &     -    &     -   &     -   &     -   &   0.024$^c$ &   0.010 \\
  203  & New$^g$ &     -    &     -   &     -   &     -   &   0.71$^c$  &   0.28  \\
  205  & New$^g$ &     -    &     -   &     -   &     -   &   0.17$^c$  &   0.067 \\
  206  & New$^g$ &     -    &     -   &     -   &     -   &   0.099$^c$ &   0.040 \\
  208  & New$^g$ &     -    &     -   &     -   &     -   &   0.058$^c$ &   0.023 \\
  210  & New$^g$ &     -    &     -   &     -   &     -   &   0.067$^c$ &   0.027 \\
  211  & New$^g$ &     -    &     -   &     -   &     -   &   0.068$^c$ &   0.027 \\

            \noalign{\smallskip}
            \hline
         \end{tabular}
      \]
{The sample of 21 Class~II sources with $-0.05 \leq \aircl \leq 0.55$
might contain a significant population of transition objects (flat-spectrum
objects) between 
Class~I protostars and Class~II T~Tauri stars.}
\begin{list}{}{}
\item[ $^{a}$ ]{The ISO number refers to the numbering in Table~1 (available 
only in electronic form at http://cdsweb.u-strasbg.fr/).}
\item[ $^{b}$ ]{The J-band flux of ISO127$\,=\,$GY245 is taken from Greene et al. 
(\cite{gwayl}).}
\item[ $^{c}$ ]{ISO77$\,=\,$GY152, ISO85$\,=\,$CRBR51, and the last 13 sources
are not detected in all near-IR bands, and
no reliable $M_J$ or $M_H$ can be estimated. $\ls$ has been obtained using 
Eq.~8 (Sect.~4.2).}
\item[ $^{d}$ ]{ISO204$\,=\,$L1689-IRS5 and ISO212$\,=\,$L1689-IRS7 refer to 
the IR sources listed by Greene et al. (\cite{gwayl}) in L1689.}
\item[ $^{e}$ ]{ISO196$\,=\,$WSB60 corresponds to the source B162816$-$243657 
in Barsony et al. (\cite{bklt}).}
\item[ $^{f}$ ]{Since ISO90$\,=\,$WL22 and ISO92$\,=\,$WL16 are young early-type
stars (see Sect.~3.1), $\ls$ cannot be derived accurately using the method 
described in Sect.~4.1. The quoted luminosities for WL22 and WL16 
are taken from WLY89 and Comer\'on et al. (\cite{crbr}), respectively.
They have been scaled to $d=140\,$pc.}
\item[ $^{g}$ ]{The adopted names and the J2000 coordinates of the
completely new IR sources are given in Table~1.}
\end{list} 
   \end{table*}


   \begin{table*}
      \caption[]{Class~III YSOs}
         \label{class3}
      \[
         \begin{tabular}{clccccc}
            \hline
            \noalign{\smallskip}
ISO     & Identification &  $\aircl$  & $M_J$ & $M_H$ & $\av$ &  $\ls$ \\
\#$^a$  &                & ($\airc$)  & [mag] & [mag] & [mag] & [$\sol$] \\
            \noalign{\smallskip}
            \hline
            \noalign{\smallskip}
	    
  114  & WL19/GY227$^b$     &  $-$0.05  &     -   &  $-$1.0 &  73.4 &  52.    \\
  125  & WL5/GY246$^b$      &  $-$1.02  &     -   &  $-$0.7 &  59.8 &  39.    \\
   34  & GY12$^b$           &  $-$1.06  &   3.3 &     -   &  19.5 &   0.89  \\
   58  & WL8/GY96$^b$       &  $-$1.11  &   1.9 &     -   &  35.5 &   4.2   \\
  152  & GY289$^b$          &  $-$1.33  &   3.3 &     -   &  27.5 &   0.90  \\
  198  & SR20           &  $-$1.65  &   1.9 &     -   &   4.0 &   4.0   \\
  133  & GY253$^b$          &  $-$1.73  &   3.3 &     -   &  31.2 &   0.85  \\
   45  & LFAM8/SKS1$-$19$^b$  &  $-$1.82  &   4.2 &     -   &  25.2 &   0.34  \\
   80  & GY156$^b$          &  $-$2.00  &   3.4 &     -   &  22.5 &   0.84  \\
   10  & DoAr21/GSS23$^b$   &  $-$2.00  &   0.7 &     -   &   6.0 &  15.    \\
   27  & WSB28          &  $-$2.06  &   4.1 &     -   &   4.3 &   0.36  \\
  149  & B162730$-$244726 &  $-$2.20  &   3.7 &     -   &  10.3 &   0.61  \\
  135  & VSSG22         &  $-$2.23  &   3.0 &     -   &  17.1 &   1.2   \\
   64  & VSSG11$^b$         &  $-$2.24  &   3.4 &     -   &  15.2 &   0.83  \\
   73  & VSSG3/GY135    &  $-$2.26  &   2.3 &     -   &  15.7 &   2.8   \\
  180  & VSSG14/GY372   &  $-$2.51  &   2.2 &     -   &   5.5 &   3.0   \\
  184  & IRS55/GY380    &  $-$2.65  &   2.6 &     -   &   6.0 &   1.8   \\
   11  & VSSG19$^b$         &  $-$2.75  &   3.8 &     -   &   3.9 &   0.50  \\
    7  & GSS20          &  $-$2.84  &   3.1 &     -   &   4.7 &   1.1   \\
   60  & GY101$^b$          & ( 0.00) &     -   &   2.0 &  55.5 &   2.0   \\
   61  & GY103$^b$          & ($-$0.45) &     -   &   2.6 &  48.5 &   1.1   \\
  126  & GY248$^b$          & ($-$0.82) &   4.2 &     -   &  25.5 &   0.36  \\
  101  & IRS30/GY203$^b$    & ($-$0.99) &     -   &   2.2 &  36.5 &   1.6   \\
   14  & B162607$-$242742 & ($-$1.43) &   3.5 &     -   &  20.6 &   0.70  \\
    8  & B162601$-$242945 & ($-$1.96) &   3.9 &     -   &   8.1 &   0.47  \\
  183  & GY377$^b$          & ($-$1.98) &   4.0 &     -   &  16.0 &   0.40  \\
  157  & GY296$^b$          & ($-$2.02) &   5.5 &     -   &   5.1 &   0.081 \\
    4  & B162541$-$242138 & ($-$2.05) &   5.8 &     -   &   6.7 &   0.060 \\
   69  & GY122          & ($-$2.17) &   5.5 &     -   &   2.7 &   0.080 \\
   96  & GY193          & ($-$2.24) &   4.2 &     -   &   7.4 &   0.34  \\
   97  & GY194          & ($-$2.35) &   4.1 &     -   &   9.1 &   0.40  \\
  188  & GY410          & ($-$2.36) &   4.0 &     -   &  10.2 &   0.43  \\
   66  & GY112          & ($-$2.42) &   4.2 &     -   &   3.6 &   0.35  \\
  130  & SR12/GY250$^b$     & ($-$2.48) &   3.4 &     -   &   1.2 &   0.83  \\
    5  & IRS10          & ($-$2.78) &   2.7 &     -   &   5.4 &   1.7   \\
   -   & IRS50/GY306    &    $-$    &   3.7 &     -   &  11.5 &   0.60  \\
   16  & SR3/GSS25      &    $-$    &   -   &     -   &     -   & 100.$^c$ \\
   48  & S1/GY70$^b$        &    $-$    &   -   &     -   &     -   & 1100.$^c$ \\

            \noalign{\smallskip}
            \hline
         \end{tabular}
      \]
\begin{list}{}{}
\item[ $^a$ ]{The ISO number refers to the numbering in Table~1 (available 
only in electronic form at http://cdsweb.u-strasbg.fr/).}
\item[ $^b$ ]{Class~III YSOs located inside the CS contours of Fig.~1 (see
Sect.~3.5 and 4.4).}
\item[ $^c$ ]{For the two B stars ISO16$\,=\,$SR3 and ISO48$\,=\,$S1 (e.g. Elias \cite{elias}),  
$\ls$ cannot be derived using the method described in Sect.~4.1. The quoted
values are taken from Lada \& Wilking (\cite{lw84}), 
and have been scaled to $d=140\,$pc.}
\end{list} 
   \end{table*}

   \begin{table*}
      \caption[]{Class~III candidates located within the CS contours 
of Fig.~1}
         \label{cl3cand}
      \[
         \begin{tabular}{clccccc}
            \hline
            \noalign{\smallskip}
ISO     & Identification &  $\aircl$ & $M_J$ & $M_H$ & $\av$ &  $\ls$  \\
\#$^a$  &                & ($\airc$) & [mag] & [mag] & [mag] & [$\sol$] \\
            \noalign{\smallskip}
            \hline
            \noalign{\smallskip}

  191  & GY463          &  $-$1.46  &   4.4 &     -   &  25.2 &   0.28  \\
   44  & B162628$-$241543 &  $-$1.63  &   3.9 &     -   &  21.7 &   0.45  \\
   81  & VSSG7/GY157    &  $-$2.13  &   2.1 &     -   &  29.8 &   3.4   \\
   91  & VSSG8/GY181    &  $-$2.20  &   2.4 &     -   &  22.9 &   2.4   \\
   28  & B162621$-$241544 &  $-$2.23  &   2.9 &     -   &  15.5 &   1.4   \\
   18  & SKS1$-$7         &  $-$2.38  &   3.6 &     -   &  17.4 &   0.64  \\
   25  & CRBR17         & ( 0.68) &   8.3 &     -   &  10.1 &   0.004 \\
  153  & GY290          & ( 0.41) &     -   &   3.7 &  41.5 &   0.31  \\
  100  & B162705$-$244013$^b$ & ( 0.12) &     -   &     -   &     -   &     -   \\
  109  & GY215$^b$          & ( 0.05) &     -   &     -   &     -   &     -   \\
  104  & GY207          & ($-$0.31) &     -   &   4.6 &  39.5 &   0.11  \\
   49  & B162636$-$241811$^b$ & ($-$0.36) &     -   &     -   &     -   &     -   \\
   15  & CRBR4          & ($-$0.41) &     -   &   1.9 &  58.9 &   2.2   \\
  146  & GY278          & ($-$0.58) &     -   &   2.4 &  47.8 &   1.3   \\
  136  & GY258          & ($-$0.72) &   8.2 &     -   &  10.8 &   0.004 \\
   55  & IRS16/GY92     & ($-$0.80) &     -   &   5.0 &  29.1 &   0.079 \\
  173  & IRS53/GY334    & ($-$0.81) &     -   &   2.0 &  40.2 &   2.1   \\
   71  & GY130          & ($-$0.87) &     -   &   3.7 &  38.8 &   0.31  \\
  162  & GY309          & ($-$1.01) &     -   &   3.3 &  42.5 &   0.47  \\
   50  & B162636$-$241902 & ($-$1.06) &   6.6 &     -   &  14.0 &   0.025 \\
   57  & B162641$-$241801 & ($-$1.22) &     -   &   3.1 &  32.3 &   0.58  \\
  111  & WL9/GY220      & ($-$1.27) &   5.5 &     -   &  21.6 &   0.086 \\
   22  & B162618$-$241712 & ($-$1.50) &   4.2 &     -   &  26.8 &   0.34  \\
  113  & IRS32/GY228    & ($-$1.52) &   3.6 &     -   &  18.5 &   0.62  \\
   82  & GY163          & ($-$1.53) &     -   &   4.1 &  32.8 &   0.19  \\
  131  & GY255          & ($-$1.57) &   4.2 &     -   &  21.6 &   0.35  \\
  169  & GY322          & ($-$1.59) &   4.0 &     -   &  16.3 &   0.43  \\
   47  & IRS14/GY54     & ($-$1.85) &   4.7 &     -   &  16.6 &   0.21  \\
  189  & GY412          & ($-$1.90) &   5.2 &     -   &  17.6 &   0.12  \\
  181  & GY373          & ($-$1.91) &   6.4 &     -   &   7.7 &   0.032 \\
  174  & GY346          & ($-$1.93) &   5.6 &     -   &  19.5 &   0.075 \\
   74  & IRS20/GY143    & ($-$1.97) &   3.8 &     -   &  16.0 &   0.50  \\
  179  & GY370          & ($-$1.99) &   5.4 &     -   &  10.0 &   0.094 \\
  148  & GY283          & ($-$2.09) &   4.3 &     -   &  13.1 &   0.32  \\
   42  & VSSG29/GY37    & ($-$2.14) &   6.8 &     -   &   6.5 &   0.021 \\
  192  & GY472          & ($-$2.14) &   5.9 &     -   &  20.1 &   0.054 \\
  186  & GY398          & ($-$2.20) &   5.2 &     -   &   5.2 &   0.11  \\
  158  & GY297          & ($-$2.22) &   6.0 &     -   &   0.7 &   0.047 \\
  156  & GY295          & ($-$2.52) &   4.3 &     -   &   4.8 &   0.32  \\

            \noalign{\smallskip}
            \hline
         \end{tabular}
      \]
\begin{list}{}{}
\item[ $^a$ ]{The ISO number refers to the numbering in Table~1 (available 
only in electronic form at http://cdsweb.u-strasbg.fr/).}
\item[ $^b$ ]{$\av$ and $\ls$ were not derived for these 
sources as they were detected only in the K band by Barsony et al. 
(\cite{bklt}).}
\end{list} 
   \end{table*}

\end{document}